\documentclass[
aps,prd,
twocolumn,
superscriptaddress,nofootinbib,longbibliography]{revtex4-2}
\usepackage{cmap}
\usepackage[utf8]{inputenc}
\usepackage[T1]{fontenc}
\usepackage[english]{babel}
\usepackage{amsmath,amsfonts,amssymb,mathtools}
\usepackage{nccmath}
\usepackage[width=18cm,height=24cm]{geometry}
\usepackage[dvipsnames]{xcolor}
\usepackage{hyperref}
\hypersetup{
	colorlinks,%
	citecolor=RoyalPurple,%
	filecolor=blue,%
	linkcolor=ForestGreen,%
	urlcolor=WildStrawberry
}
\usepackage{indentfirst}

\makeatletter
\def\l@subsubsection#1#2{}
\makeatother

\numberwithin{equation}{section}

\DeclareMathOperator{\Tr}{Tr}
\DeclareMathOperator{\tr}{tr}

\newcommand{\bk}{\boldsymbol{k}}
\newcommand{\bp}{\boldsymbol{p}}

\newcommand{\bx}{\boldsymbol{x}}
\newcommand{\bpar}{\boldsymbol{\partial}}
\newcommand{\tdt}{\tau\partial_\tau}
\newcommand{\bR}{\boldsymbol{R}}

\begin{document}
	
	\title{Non-vacuum gravitational effective action}
	\author{Andrei O. Barvinsky}
	\email{barvin@td.lpi.ru}
	\affiliation{Theory Department, Lebedev Physics Institute, Leninsky Prospect 53, Moscow 119991, Russia}
	\affiliation{Institute for Theoretical and Mathematical Physics, Moscow State University, 119991, Leninskie Gory, GSP-1, Moscow, Russia}
	\author{Farahmand Hasanov}
	\email{khasanov.mkh@phystech.edu}
	\affiliation{Moscow Institute of Physics and Technology, 141700, Institutskiy pereulok, 9, Dolgoprudny, Russia}
	\author{Nikita Kolganov}
	\email{nikita.kolganov@phystech.edu}
	\affiliation{Theory Department, Lebedev Physics Institute, Leninsky Prospect 53, Moscow 119991, Russia}
	\affiliation{Moscow Institute of Physics and Technology, 141700, Institutskiy pereulok, 9, Dolgoprudny, Russia}
	
	\begin{abstract}\small
		Curvature expansion for the heat kernel trace and the one-loop effective action is built for the wave operator of the theory in the quasi-thermal setup of a nonvacuum quantum state. This setup implies a non-static and non-stationary Euclidean gravitational background with periodic boundary conditions of the period $\beta=1/T$, where $T$ plays the role of effective global temperature to be locally rescaled by the metric gravitational potential. The results are obtained in the approximation quadratic in metric perturbations on top of flat Euclidean space and covariantized in terms of spacetime curvature. Covariantization includes a special vector field $\xi^\mu(x)$ which generalizes the Killing vector of static geometries with time translation isometry to the case of a generic arbitrarily inhomogeneous metric subject to timelike periodicity condition. This vector field is obtained as a covariant metric functional to quadratic order in metric perturbations and gives rise to the local function $T/\sqrt{\xi^2(x)}$, $\xi^2(x)=g_{\mu\nu}(x)\xi^\mu(x)\xi^\nu(x)$, reducing to Tolman temperature $T/\sqrt{g_{00}(\boldsymbol{x})}$ on stationary manifolds with Killing symmetry. High ``temperature'' asymptotic behavior of the nonlocal form factors---operator coefficients of the curvature tensor structures in the heat kernel and effective action---are obtained and possible cosmological applications of these results are discussed.
	\end{abstract}

	\maketitle
	


\section{Introduction}
	The purpose of this work is to extend the results of \cite{Barvinsky:1987uw,Barvinsky:1990up} on the covariant expansion of gravitational effective action to non-vacuum states potentially applicable in a wide range of non-equilibrium and nonlocal phenomena. The underlying curvature expansion was developed in the form of local Schwinger-DeWitt or Gilkey-Seeley heat kernel technique \cite{Schwinger:1951nm,DeWitt:1964mxt,Barvinsky:1985an,Seeley:1967ea,Gilkey:1975iq} and its nonlocal version pioneered in~\cite{Deser:1976yx} and later promoted to a universal method for generic field models in curved spacetime \cite{Barvinsky:1987uw,Barvinsky:1990up,Barvinsky:1993en}. It has proved to be very efficient in calculation of divergences, anomalies and nonlocal effects in massless theories, and also underlies computations in such challenging problems, like renormalization of Ho\v{r}ava gravity models \cite{Barvinsky:2021ubv}, where computational attempts based on conventional Feynman diagrammatic methods fail or the endeavors to proceed simply fear to tread.
	
	A serious limitation of this nonlocal expansion for the ultraviolet (UV) finite part of the effective action, which is sensitive to the global spacetime geometry, is that in \cite{Barvinsky:1987uw,Barvinsky:1990up} it was developed only for vacuum quantum states in asymptotically flat spacetime. Apart from several sound applications in cosmology and quantum field theory like \cite{Donoghue:2014yha,Donoghue:2015nba,Donoghue:2017pgk}, it was very often inappropriately extrapolated to black hole systems with a horizon, spatially closed models or the models with (A)dS-type boundary conditions, etc.\ (see some references in \cite{Donoghue:2017pgk}). Moreover, it does not include modifications due to the choice of a nontrivial quantum state encoded either as quantum initial data (in-in problem \cite{Barvinsky:1987uw,Adshead:2009cb}) or as in-out states of the physical scattering theory setup. Here we try to fill up this omission for one-loop effective action within covariant expansion up to the second order in spacetime curvature and quantum matter field strengths.
	
	Modelling a generic quantum state of the physical system is, of course, subject to a vast functional ambiguity in the choice of its (pure or mixed) density matrix. As is well known, one of the simplest cases of specification of this ambiguity is the situation with equilibrium canonical ensemble in static external fields, characterized by a single temperature parameter $T=1/\beta$ and canonical density matrix whose partition function is represented by the Euclidean path integral over periodic fields with the period $\beta$.

In fact, this setup can be generalized to non-equilibrium non-stationary states with their density matrix also prescribed by the Euclidean path integral, while the relevant generating functional for evolving quantum correlators is formulated in terms of quantum fields periodic on the Schwinger-Keldysh closed contour in the complex plane of time \cite{Schwinger:1960qe,Keldysh:1964ud,Jordan:1986ug,Calzetta:1986ey,Ford:2004wc,Onemli:2002hr,Adshead:2009cb}. Within this approach the generating functional can be obtained by analytic continuation from the real Euclidean time segment of length $\beta$ \cite{Barvinsky:2023jkl}, while the systems with a generic density matrix would follow from it by the variation of its sources. Thus, within this line of reasoning, specification of a rather generic quantum state consists in the calculation of effective action for quantum fields periodic on the complex Schwinger-Keldysh contour. The first step of this program is the effective action calculation on a real Euclidean time segment of $\beta$ length, which can be called a non-stationary quasi-thermal setup.
	
	Local curvature expansion of the one-loop effective action at finite temperature for {\em static} geometries of spacetime metric with vanishing shift functions $g_{0i}=0$ was built in \cite{Dowker:1988jw}. Non-local expansion for static geometries was developed in \cite{Gusev:1998rp} and subsequently extended to the stationary (non-zero $g_{0i}$) case in \cite{Elias:2017wkr,Valle:2025fev}. It leads to the thermal generalization of the Schwinger-DeWitt expansion, featuring the high temperature limit in terms of the local Tolman temperature parameter $T_{\text{Tolman}}(\bx)$ related to the Euclidean interval $\beta=1/T$ by the relation $T_{\text{ Tolman}}(\bx)=T/\sqrt{g_{00}(\bx)}$. Here we extend this result to geometries without timelike Killing symmetry for the main building block of quantum effective action---the relevant heat kernel of the field theory wave operator---and the effective action itself in the approximation quadratic in spacetime curvature and matter field strengths.
	
	The result turns out to be significantly more complicated due to non-stationary nature of the metric background. Moreover, it reveals a challenging problem, namely the covariantization of the final result. This involves converting the expression into a form that is manifestly covariant with respect to local diffeomorphisms and local gauge transformations in the matter sector, specifically, a form constructed solely in terms of spacetime and fibre-bundle curvature tensors and their covariant derivatives. This problem (called later in \cite{Donoghue:2015xla} the nonlinear completion) was encountered in \cite{Barvinsky:1987uw,Barvinsky:1990up} in the course of derivations for the vacuum case, where it was solved by the following procedure.
	
	First, the spacetime metric gets decomposed into the sum of flat space metric $\tilde g_{\mu\nu}$ and the perturbation $h_{\mu\nu}$, $g_{\mu\nu}=\tilde g_{\mu\nu}+h_{\mu\nu}$, but this split is performed in a generic coordinate system, rather than the one chosen by the Cartesian coordinate gauge $\tilde g_{\mu\nu}=\eta_{\mu\nu}$. In a generic coordinate system the definition of $\tilde g_{\mu\nu}$ is determined solely by the condition of its vanishing Riemann curvature $R^\alpha_{\;\;\beta\mu\nu}(\tilde g)=0$ which can be rewritten as the equation for the metric perturbation $h=h_{\mu\nu}$, $R^\alpha_{\;\;\beta\mu\nu}(g-h)=0$. This equation can iteratively be solved for $h$ in some preferably Lorentz-invariant gauge imposed on the unknown $h$, which yields this perturbation as a nonlocal covariant series in the Riemann (or Ricci) curvature. For example, in the DeWitt background covariant gauge, $\tilde{\nabla}^\nu h_{\mu\nu}-\tfrac12\tilde{\nabla}_\mu h=0$, this covariant series begins with the nonlocal term
	$h_{\mu\nu}=\tfrac2\Box R_{\mu\nu}+\mathcal{O}(R_{\alpha\beta\gamma\delta}^2)$,
	where $\tfrac1\Box$ denotes the Green's function of the covariant d'Alembertian acting on second rank Ricci tensor $R_{\mu\nu}=R^\alpha_{\;\;\mu\alpha\nu}$.
	
	Second, the perturbation theory in $h_{\mu\nu}$ is developed for the quantity in question (that is the heat kernel of the wave operator or the corresponding effective action---the functional determinant of the wave operator). On top of the flat space background metric this step does not differ much from a conventional Feynman diagrammatic method whether it is performed in the coordinate or momentum space representation (see, for example, \cite{Codello:2012kq}). The resulting Lorentz-invariant series in $h_{\mu\nu}$ then was converted into the series in the original curvature tensor by substituting the expression for $h_{\mu\nu}$ in terms of $R^\alpha_{\;\;\beta\mu\nu}(g)$ and by rewriting all partial derivatives and flat space metric coefficients, occurring in the operator coefficients of $h_{\mu\nu}$, in terms of covariant derivatives $\nabla_\mu$ and the full metric tensor $g_{\mu\nu}$. The same steps can be undertaken in the matter sector of the theory, where the role of $h_{\mu\nu}$ and $R^\alpha_{\;\;\beta\mu\nu}$ is played by the connection and fibre bundle curvature.
	
	Ultimately, this procedure, carried out order by order in powers of the curvature yields manifestly Lorentz-covariant and manifestly diffeomorphism invariant answer for the heat kernel, effective action and other objects in question. For the theory with the wave operator $F$,
	\begin{equation}
		F = -\Box + \hat P, \label{F}
	\end{equation}
	this expansion procedure looks as follows. Here the covariant d'Alembertian $\Box=g^{\mu\nu}\nabla_\mu\nabla_\nu$ and metric-preserving covariant derivative $\nabla_\mu$ are acting on generic fibre bundle fields in the Euclidean $D=d+1$ dimensional spacetime, $x^\mu=(x^0,\bx^i)$, $i=1,\ldots d$, while $\hat P$ is some potential term. Hats denote matrix-valued objects in the vector space of fields, $\hat 1$ in what follows denotes the unit matrix and $\tr$ implies the matrix trace. Generically, this operator is characterized by the set of Riemann $R^\alpha_{\;\;\beta\mu\nu}$, Ricci $R_{\mu\nu}$, scalar $R$ curvatures in the gravitational sector and field strengths in the sector of matter fields fibre bundle curvature $\hat{\cal R}_{\mu\nu}\equiv[\nabla_\mu,\nabla_\nu]$, along with the potential ``curvature'' $\hat P$.
	
{In this generic setup the heat kernel of this operator---the kernel of its operator exponent with the proper time parameter $\tau$ \cite{Barvinsky:1985an,Vassilevich:2003xt}---and its functional trace,
	\begin{eqnarray}
		&&K(\tau)=\hat K(\tau|x,x') = e^{-\tau F}\delta(x,x'),  \label{heat_kernel}\\
		&&\Tr K(\tau)=\int d^Dx\,\tr\hat K(\tau|x,x),  \label{TrK}
	\end{eqnarray}
serve for the calculation of the one-loop effective action}
	\begin{equation}
		W = \frac12 \Tr\ln F=-\frac12\int_0^\infty\frac{d\tau}\tau\,\Tr K(\tau).  \label{effective_action}
	\end{equation}
	For vacuum states the functional trace $\Tr$ is determined over the functional space of square integrable functions with zero boundary conditions at infinity of the Euclidean spacetime, and the range of integration over all coordinates is from $-\infty$ to $+\infty$.

	Covariant curvature expansion derived for such a setup in \cite{Barvinsky:1987uw,Barvinsky:1990up} has the following form up to terms quadratic in the full set of curvatures $\Re=(R^\alpha_{\;\;\beta\mu\nu},\hat{\cal R}_{\mu\nu},\hat P)$ inclusive
	\begin{widetext}
	\begin{multline} \label{eq:intro_vacuum}
		\Tr K_{\mathrm{vac}}(\tau) = \frac1{(4\pi\tau)^{D/2}}\int d^D x\,\sqrt{g}\,\tr \biggl[\hat 1+
		\tau \Bigl(\frac16 R\, \hat 1 - \hat P\Bigr)+\frac12\tau^2\Bigl(R_{\mu\nu} \, f^{\mathrm{vac}}_1(-\tau\Box) \, R^{\mu\nu}\hat 1
		\\+ R \, f^{\mathrm{vac}}_2(-\tau\Box) \, R\hat 1
		+ \hat P \, f^{\mathrm{vac}}_3(-\tau\Box) \, R + \hat P \, f^{\mathrm{vac}}_4(-\tau\Box) \,\hat P
		+\hat{\cal R}_{\mu\nu} \, f^{\mathrm{vac}}_{\mathcal{R}}(-\tau\Box) \,\hat{\mathcal {R}}^{\mu\nu}\Bigr)\biggr]+\mathcal{O}[\Re^3],
	\end{multline}
	\end{widetext}
	where $f^{\mathrm{vac}}_I(-\tau\Box)$, $I=1,\ldots 4, \mathcal{R}$, is the set of special nonlocal operator functions of the covariant d'Alembertians acting on relevant curvatures. Obviously, this expression is fully Lorentz and diffeomorphism invariant, its nonlocal operator coefficients---functions of the full curved space d'Alembertians---are symmetric in the sense that their operator arguments can be integrated by parts without extra surface terms, which is guaranteed by the asymptotic behavior of curvatures at spacetime infinity.

	For the generalization of (\ref{eq:intro_vacuum}) to the non-vacuum case we will, basically, adopt the same strategy consisting of the perturbation theory in $h_{\mu\nu}$ and its covariantization. As a result, the underlying periodicity in one spacetime coordinate, which will now play a distinguished role as compared to the rest of coordinates, will break manifest Lorentz invariance and indicate the presence of structures additional to metric and curvature tensors. Technically this leads, along with the obvious replacement $\int_{R^D} d^D x\mapsto\int_0^\beta d x^{0} \int_{R^d} d^d\bx$, to the appearance of Lorentz noninvariant tensor terms with uncontracted time indices arising at the stage of perturbation theory in $h_{\mu\nu}$. In the case of a thermal state on static spacetime \cite{Dowker:1988jw,Gusev:1998rp,Elias:2017wkr} this additional structure is nothing but the timelike Killing vector $\bar{\xi}^\mu$ in terms of which the results acquire a manifestly diffeomorphism invariant form. Covariantization in this case proceeds by replacing any noncovariant zeroth component of any tensor by its contraction with this Killing vector field, which has a trivial form $\bar{\xi}^\mu=\delta^\mu_0$ only in the distinguished coordinate system associated with the time independent metric coefficients. In our setup such vector is a priori absent because spacetime does not possess infinitesimal Killing symmetries---in the covering space it has only finite shift ``symmetry'' associated with the displacement by the period $\beta$ in the time direction.
	
	Thus, one of the major goals of this work will be the construction of a special vector field $\xi^\mu$ generalizing the notion of the Killing vector $\bar{\xi}^\mu$ to non-stationary geometries, so that we can express the curvature expansion in a form manifestly invariant under local diffeomorphisms. This vector field will be obtained as a nonlocal covariant functional of the metric expanded to second order in metric perturbations. As we will see, this functional has a two-parameter ambiguity which can be used to simplify the formalism, but thus far we do not have good principles to fix this ambiguity.
	
	The paper is organized as follows. In Sect.~\ref{sec:heat_kernel_h} we build perturbation theory in $h_{\mu\nu}$. For simplicity we consider the case of a quantum scalar field which allows us to disregard hatted matrix structure of the formalism along with the contribution of the fibre bundle curvature $\hat{\cal R}_{\mu\nu}$. The generalization to fields with nontrivial spin-tensor structure is straightforward and will be published separately. Sect.~\ref{sec:heat_kernel_cov} is devoted to the covariantization of the result in terms of a special construction of the quasi-Killing vector field. In Sect.~\ref{sec:effective_action} we provide a brief overview of the properties and various representations of the nonlocal form factors of the obtained heat kernel and effective action, along with their high and low ``temperature'' limits.
	Sect.~\ref{sec:discussion} contains discussion of the further prospects of this work and its planned applications in the problem of quantum initial conditions for inflationary cosmology. Very involved technical details of the obtained results and their derivation will be published in the forthcoming extended paper~\cite{tbp}.
	
\section{Heat kernel perturbation theory} \label{sec:heat_kernel_h}
	In this section we calculate the heat kernel trace (\ref{TrK}) of the scalar field operator (\ref{F}) at the second order in metric perturbation $h_{\mu\nu}$ on top of the background spacetime $S^1 \times R^d$. Though the latter is locally flat, non-trivial effects arise due to periodicity in the coordinate $x^0 \in [0, \beta)$ parameterizing $S^1$ circle. This setup is associated with the periodic Euclidean time, and the corresponding period is naturally interpreted as the inverse temperature $\beta$.
		
	Assigning the potential $P$ to the linear perturbation order, we expand the operator $F = -\Box + P$ on the flat space background as $F = \sum_n F_n$, where the lower order terms $F_n=\mathcal{O}[h^n]$ have the form {(all derivatives outside of the bracketed expressions are acting fully to the right)},
		\begin{equation} \label{eq:F_comp_grav}
			\begin{aligned}
				&F_0 = -\partial_\mu^2, \qquad
				F_1 = \partial_\mu \, h^{\mu\nu} \partial_\nu - \frac12 (\partial^\mu h) \partial_\mu + P, \\
				&F_2 = -\partial_\mu \, h^\mu_\rho \, h^{\rho\nu} \, \partial_\nu + {\frac12 (h^{\rho\sigma}\partial^\mu h_{\rho\sigma})} \partial_\mu + \frac12(\partial_\mu h)\,h^{\mu \nu}\, \partial_\nu,
			\end{aligned}
		\end{equation}
and obtain the corresponding expansion for the heat kernel trace $\Tr K(\tau) = \sum_n\Tr K_n(\tau)$. The relevant individual terms have the following form
		\begin{subequations}
			\begin{align}
				&\Tr K_0(\tau) = \Tr e^{-\tau F_0}, \\[1ex]
				&\Tr K_1(\tau) = - \tau \Tr \bigl(e^{-\tau F_0} F_1 \bigr), \\[1ex]
				&\Tr K_2(\tau) =	- \tau \Tr \bigl(e^{-\tau F_0} F_2 \bigr)\nonumber\\ &\qquad+ \frac12 \tau^2 \int_0^1 d\alpha \, \Tr \bigl(e^{-\alpha \tau F_0} \, F_1 \, e^{-(1-\alpha)\tau F_0} \, F_1\bigr).
			\end{align}
		\end{subequations}
	
	Explicit expressions are obtained in the momentum-space representation with the basis eigenvectors parameterized by the spacetime momentum $p_\mu = (\omega_n, \bp)$, where $\omega_n = 2\pi n / \beta$ is the discrete Matsubara frequency and $\bp$ is the $d$-dimensional continuous spatial vector. We write the result in the following form
		\begin{subequations} \label{eq:TrK012}
			\begin{align}
				&\Tr K_0(\tau) = \frac1{(4\pi\tau)^{D/2}} \!\int \!d^D x \, r(\tau), \label{eq:TrK0}\\
				&\Tr K_1(\tau) = \frac1{(4\pi\tau)^{D/2}} \!\int \!d^D x \, \Bigl(T_{\mu\nu}(\tau) h_{\mu\nu} (x) + T(\tau)  P(x)\Bigr), \label{eq:TrK1}\\
				&\begin{aligned}
				&{}\Tr K_2(\tau) = \frac1{(4\pi\tau)^{D/2}} \frac12 \!\int \!d^D  x \,\Bigl(  h_{\mu\nu}(x)  \varPi_{\mu\nu,\rho\sigma}(\tau, \partial)  h_{\rho\sigma}(x) \\ &\hspace{4em}+ 2 P(x) \, \varPi_{\mu\nu}(\tau, \partial) \, h_{\mu\nu}(x) + P(x) \, \varPi(\tau,\partial) \, P(x) \Bigr).
				\end{aligned}\label{eq:TrK2}
			\end{align}
		\end{subequations}
	Here all integrations are carried out over $S^1 \times R^d$.

	All the coefficients of $h_{\mu\nu}$ and $P$ are linear combinations of the basic thermal function $r(\tau)$ and basic nonlocal form factor $f(\tau, \partial)$
		\begin{align}
			&r(\tau) = \frac{\sqrt{4\pi\tau}}{\beta}\sum_n e^{-\omega_n^2 \tau},  \label{eq:r_def}\\
			&f(\tau, \partial)\,{=}\!\int_0^1 \!d\alpha\, e^{\alpha(1-\alpha)\tau \partial_\mu^2}\,
	    \frac{\sqrt{4\pi\tau}}{\beta} \sum_{n} e^{-\tau (\omega_n+i\alpha\partial_0)^2}\!\!,  \label{eq:f_def}
		\end{align}
	and a finite number of their $\tau$-derivatives. The coefficients $T_{\mu\nu}(\tau)$ and $T(\tau)$ in the first order of perturbation theory are local, i.e.\ they do not contain $\partial_\mu$ and are expressed via the thermal function $r(\tau)$
	\begin{equation}
		\begin{aligned}
			&T_{\mu\nu}(\tau) = \frac12 r(\tau) \, \eta_{\mu\nu} - \tdt r(\tau) \, \bar{\xi}_\mu \bar{\xi}_\nu , \\
			&T(\tau) = - \tau \, r(\tau),
		\end{aligned}
		\label{eq:TrK1_Tdef}\\
	\end{equation}
	while the rest of the coefficients are nonlocal operators involving $f(\tau, \partial)$
		\begin{equation} \label{eq:Pi_ansatz}
		 	\begin{aligned}
		 		&\varPi_{\mu\nu,\rho\sigma}(\tau, \partial) =
		 		\sum_{a=1}^{14} C_a(\tau, \partial) \, (\varPi_a)_{\mu\nu,\rho\sigma}(\partial), \\
		 		&\varPi_{\mu\nu}(\tau, \partial) =  \sum_{a=15}^{18} C_a(\tau, \partial) \, (\varPi_a)_{\mu\nu}(\partial),\\[1ex]
		 		&\varPi(\tau, \partial) =  \tau^2 \, f(\tau, \partial)
		 	\end{aligned}
		 \end{equation}
	Their tensor structure is determined by the flat space metric $\eta_{\mu\nu}$, partial derivative operator $\partial_\mu$ and the vector $\bar{\xi}^\mu = \delta^\mu_0$ along with appropriate symmetry properties. The origin of the vector $\bar{\xi}^\mu$, which distinguishes the Euclidean time direction in these expressions, corresponds to the breaking of $SO(D)$-invariance down to $SO(d)$-invariance due to the distinguished role of time direction. Specifically, there are 14 independent combinations $(\varPi_a)_{\mu\nu,\rho\sigma}$ constructed out of $\eta_{\mu\nu}$, $\partial_\mu$ and $\bar{\xi}_\mu$, that have symmetries $\mu \leftrightarrow \nu$, $\rho \leftrightarrow \sigma$ and $(\mu,\nu) \leftrightarrow (\rho,\sigma)$. Similarly, there are 4 combinations $(\varPi_a)_{\mu\nu}$ with the same constituents having the symmetry $\mu\leftrightarrow\nu$. These basis combinations are listed in Appendix~\ref{sec:app_basis_tens}.
	
	The operator coefficient functions $C_a$ of the ansatz (\ref{eq:Pi_ansatz}) are presented in Appendix~\ref{sec:app_cfs_ans} in the momentum space representation. They have the structure
	\begin{equation}
		\begin{aligned}C_a(\tau, \partial) = \, &\sum_{p=0}^2 \tau^p \, c^{(p)}_a(\partial,\tdt) f(\tau,\partial)
			\\[-0.5ex]+ &\sum_{p=0}^1 \tau^p \, d^{(p)}_a(\partial,\tdt)\, r(\tau),     \end{aligned}            \label{eq:cfs_fun_form}
	\end{equation}
	where $c^{(p)}$ and $d^{(p)}$ are polynomials in $\tdt$, $\partial_0$ and Laurent polynomials in $\partial_\mu^2$ and~$\bpar^2 = \partial_i \partial^i$. Eq.~(\ref{eq:cfs_fun_form}) imply grading $p$ assigned to the coefficients of the functions $r(\tau)$ and $f(\tau)$ in $C_a$ (and $\varPi$) of the quadratic part $\Tr K_2(\tau)$ according to the power $p$ of $\tau$, (with $\partial_\tau$ assigned a negative power $-1$ and the argument of $r(\tau)$ and $f(\tau, \partial)$ not participating in this grading).	Similarly the expression for the linear part $\Tr K_1(\tau)$ follows the same grading scheme, specifically $T_{\mu\nu}(\tau)$ and $T(\tau)$ in (\ref{eq:TrK1_Tdef}) have grading 0 and 1, respectively. The explicit answer for the coefficient functions $c^{(p)}(\partial,\tdt)$ and $d^{(p)}(\partial,\tdt)$ in the momentum space representation,\footnote{Hereinafter, we will use the notation $k_0$ instead of $\omega_m = 2\pi m /\beta$ noting that $k_0$ is discrete.} $\partial_\mu=ik_\mu=(ik_0,i\bk)$, $k_0\equiv\omega_m\equiv 2\pi m / \beta$ can be read off the table in Appendix~\ref{sec:app_cfs_ans} as the coefficients of $f(\tau, \partial)$ and $r(\tau)$ of relevant powers of $\tau$.

\section{Covariant form of the heat kernel trace} \label{sec:heat_kernel_cov}
	The effective action is invariant under local diffeomorphisms. However, the structure of the result does not exhibit manifest diffeomorphism invariance. This happens because local diffeomorphisms---Lie derivatives along vector field $\zeta^\nu$,
		\begin{align} \label{eq:diffeo_def}
				&\delta_\zeta h_{\mu\nu} = \partial_\mu \zeta_\nu + \partial_\nu \zeta_\mu + \zeta^\rho \partial_\rho h_{\mu\nu} + \partial_\mu \zeta^\rho h_{\rho\nu} + \partial_\nu \zeta^\rho h_{\mu\rho}, \nonumber\\
				&\delta_\zeta P = \zeta^\mu \partial_\mu P,
		\end{align}
	mix the orders of perturbation theory in $h_{\mu\nu}$, so that $\Tr K_1$ and $\Tr K_2$ are not invariant on their own, but demonstrate local invariance only together and only up to terms cubic in the perturbation. In terms of the representation (\ref{eq:TrK1}),~(\ref{eq:TrK2}) this invariance implies the following Ward (or Noether) identities, which in the Fourier representation, $\partial_\mu = i k_\mu$, read as the relations between the first and second order form factors
		\begin{align} \label{eq:Ward_ids}
			&k^\sigma \varPi_{\mu\nu,\rho\sigma}(\tau, ik) {=} \frac12 k^\sigma\bigl( \eta_{\rho\sigma} T_{\mu\nu}(\tau) {-} \eta_{\mu\rho} T_{\sigma \nu}(\tau) {-} \eta_{\rho\nu} T_{\sigma\mu}(\tau)\bigr), \nonumber\\
			&k^\sigma \varPi_{\rho\sigma}(\tau, ik) = \frac12 k_{\rho} \, T(\tau).
		\end{align}
	One can verify that these Ward identities hold for the form factors (\ref{eq:TrK1_Tdef}) and (\ref{eq:Pi_ansatz}) by direct inspection.
	
	Moreover, though the background spacetime is \emph{locally} flat, the expansion (\ref{eq:TrK012}) is not Lorentz invariant.
	The reason for the Lorentz invariance breakdown is the distinguished role of the spacetime direction associated with the periodicity in time, which reflects the \emph{global} property of the background manifold. This results in the appearance of uncontracted indices of zeroth tensor components.
	
	This loss of manifest symmetries can be circumvented by replacing the perturbation theory in terms of metric perturbation by the expansion in powers of the diffeomorphism covariant objects---Riemann and Ricci curvatures, characterizing the operator (\ref{F}), their covariant derivatives and a vector field structure reflecting periodicity in the distinguished time coordinate. This vector field $\xi^\mu(x)$, which on flat-space background $h_{\mu\nu}=0$ reduces to $\bar\xi^\mu=\delta^\mu_0$ (in the Cartesian coordinates used in the expansion (\ref{eq:TrK012})), should convert noninvariant zeroth components of various tensors into diffeomorphism invariant contractions. For instance, in the case of the Ricci tensor and covariant derivatives this conversion looks respectively like $R_{00}\mapsto R_{\mu\nu}\xi^\mu\xi^\nu$ and $\nabla_0\mapsto\xi^\mu\nabla_\mu$. For such contractions to be invariant $\xi^\mu=\xi^\mu[h]$ should be a functional of $h_{\mu\nu}$ transforming under (\ref{eq:diffeo_def}) as a vector field and should be explicitly constructed. As we will see, such a construction is indeed possible at least up to the second order in $h_{\mu\nu}$ and, moreover, it turns out to be not unique.
	
	As mentioned in Introduction, the above procedure of covariantization was systematically implemented for the vacuum effective action in asymptotically flat spacetime in \cite{Barvinsky:1987uw,Barvinsky:1990up}, where it was essentially simpler in view of the absence of the $\xi^\mu$ vector structure. Presence of this structure in non-vacuum case makes the situation much more involved and not so systematic, but still manageable in the approximation quadratic in curvatures.
	
	Briefly, the conversion procedure begins with the choice of the basis of curvature invariants of the zeroth, first and second order in the Ricci curvature\footnote{Riemann curvature does not obviously arise in the zeroth and first orders, while in the second order it can be reexpressed via integration by parts in terms of the Ricci tensor---the same feature as in the vacuum case of \cite{Barvinsky:1987uw,Barvinsky:1990up} where Riemann tensor nonlocally expresses in terms of the Ricci one to all orders of covariant perturbation theory.} and the potential $P$, the vector field $\xi^\mu$ participating in the construction of the chosen basis invariants via its contractions with the curvature indices and covariant derivatives. All curvature invariants are then taken with the unknown local or operator-valued coefficients to form a full manifestly covariant quadratic in curvatures ansatz. Then the ansatz expression is expanded in powers of the metric perturbation $h_{\mu\nu}$ and compared to the known noncovariant expansion (\ref{eq:TrK0})--(\ref{eq:TrK2}). This yields the system of equations for the unknown coefficients---local functions or nonlocal form factors---along with the vector field $\xi^\mu=\xi^\mu[h]$ which is found from these equations as a functional of $h_{\mu\nu}$.
	
	Obtaining these equations is strongly facilitated by the fact that the expansion (\ref{eq:TrK0})--(\ref{eq:TrK2}), due to the structure of its form factors (\ref{eq:cfs_fun_form}), can be split into groups of terms which are separately diffeomorphism invariant, so that their solution for unknown form factors can be found independently in each of these groups. This split follows from the intrinsic structure of the functions (\ref{eq:cfs_fun_form}) and the following observations: 1) the basic form factor $f(\tau, \partial)$ enters only $\Tr K_2$, hence the corresponding parts are invariant separately; 2) all terms in $\Tr K$ (cf.\ (\ref{eq:cfs_fun_form}) and (\ref{eq:TrK1_Tdef})) are weighted by the powers $p=0,1,2$ of $\tau$, hence the coefficient of each power $\tau^p$ is invariant separately; 3) $\Tr K_n$ contains powers $p \le n$, $n=0,1,2$.

\subsection{Invariant form in stationary case}
		
	Implementation of the above procedure is much easier on a stationary metric background when $\xi^\mu$ is a Killing vector satisfying the equation $\nabla_\mu\xi_\nu+\nabla_\nu\xi_\mu=0$, so that it is not a functional of the metric but a part of the physical setup---a fixed vector field which in a special coordinate system reads as $\xi^\mu=\delta^\mu_0$. In this coordinate system stationarity means that the metric itself and its perturbations, along with the potential $P$, are independent of the Euclidean time, i.e. $h_{\mu\nu} = h_{\mu\nu}(\bx)$ and $P=P(\bx)$. Technically, stationary case corresponds to setting $\partial_0 = 0$ in all the form factors obtained above. This, in particular, implies that $\partial_\mu^2 = \bpar^2$, so we will use the notation $\partial^2\equiv\partial_\mu^2$ in the stationary case.
	
	The main simplification due to stationarity consists in a special factorization of the basic form factor (\ref{eq:f_def}) into the product of the vacuum form factor $f_{\mathrm{vac}}(-\tau \partial^2)$, introduced in \cite{Barvinsky:1987uw}, and the basic thermal function (\ref{eq:r_def})
		\begin{eqnarray}
			&&\bar{f}(\tau, \partial) \equiv f(\tau, \partial)\Bigr|_{\partial_0=0} = f_{\mathrm{vac}}(-\tau \partial^2) \, r(\tau), \label{eq:ff_stationary}\\
			&&f_{\mathrm{vac}}(-\tau \partial^2) = \int_0^1 d\alpha \, e^{\alpha(1-\alpha)\tau \partial^2}. \label{vac_ff}
		\end{eqnarray}
	The vacuum form factor here corresponds to zero-temperature $\beta\to\infty$ limit (see separation of the form factors into vacuum and thermal parts below in (\ref{eq:f_split})). Here and in what follows the overbar denotes the stationary limit, i.e.\ the substitution of $\partial_0 = 0$ in all form factors acting on stationary background fields.
	
	The factorization property (\ref{eq:ff_stationary}) allows one to establish the structure of form factors similar to the vacuum case of \cite{Barvinsky:1987uw,Barvinsky:1990up}. In the vacuum effective action each $n$-th order of curvature expansion was rendered the form weighted by the same $n$-th power of the proper time---the property following from the dimensional arguments which keep the same dimensionality of all orders of perturbation theory (bearing in mind that the vacuum form factors are dimensionless functions of the dimensionless argument $u=-\tau\partial^2$). In view of the third of the observations above, this property was attained by subtracting the first $n$ terms of the Taylor expansion from the $n$-th order form factor and relocating them to lower orders of perturbation theory. In the non-vacuum case this is literally impossible because the thermal function $r(\tau)$ given by (\ref{eq:r_def}) is not analytic at $\tau=0$. However, the factorization (\ref{eq:ff_stationary}) makes this subtraction possible for the operator coefficient of $r(\tau)$ by introducing the notions of the subtracted vacuum form factor $f_{\mathrm{vac}}^{[p]}(-\tau\partial^2)$ and basic subtracted form factor in stationary theory $\bar f^{[p]}(\tau,\partial)$,
		\begin{eqnarray}
			&&f_{\mathrm{vac}}^{[p]}(u) = \frac1{u^p}\Bigl(f_{\mathrm{vac}}(u)
	        - \sum_{l=0}^{p-1}\frac{l!}{(2l+1)!}(-u)^l\Bigr),  \label{eq:ff_vac_subtr}\\
	        &&\bar{f}^{[p]}(\tau, \partial)
	        = f_{\mathrm{vac}}^{[p]}(-\tau \partial^2) \, r(\tau),  \label{eq:ff_subtr0}
		\end{eqnarray}
	where we imply that $f_{\mathrm{vac}}^{[0]}(u) \equiv f_{\mathrm{vac}}(u)$ and $\bar{f}^{[0]}(\tau, \partial) \equiv \bar{f}(\tau, \partial)$.
	
	This allows us to combine the first two observations above, namely isolate the part simultaneously containing the form factor $f$ and having the weight $\tau^2$. 	Expressing $\tau^p \bar{f}(\tau, \partial)$ in terms of $\tau^2 \bar{f}^{[2-p]}(\tau, \partial) = \tau^p (-\partial^2)^{p-2}\bar{f}(\tau, \partial) + \ldots$, where extra terms are proportional to $r(\tau)$, we can rewrite the general form of the coefficients (\ref{eq:cfs_fun_form}) of the ansatz (\ref{eq:Pi_ansatz}) as
	\begin{align} \label{eq:cfs_fun_form_stationary}
		\bar{C}(\tau, \partial) = \tau^2 \sum_{p=0}^2 \bar{c}^{[p]}(\partial,\tdt) \,
	    \bar{f}^{[p]}(\tau,\partial) \nonumber \\+ \sum_{p=0}^1 \tau^p \,\bar{d}^{[p]}(\partial,\tdt) \, r(\tau),
	\end{align}
	where the new coefficients $\bar{c}^{[p]}$ and $\bar{d}^{[p]}$ read in terms of the old ones as
		\begin{equation} \label{eq:cfs_subtr}
			\begin{aligned}
				&\bar{c}^{[p]}(\partial,\tdt) = (-\partial^2)^p \, \bar{c}^{(2-p)}(\partial,\tdt+p), \\[2ex]
				&\bar{d}^{[p]}(\partial, \tdt) = \bar{d}^{(p)}(\partial,\tdt)
	            \\ &\hspace{3em}+ \sum_{l=0}^p\frac{l!}{(2l+1)!} (\partial^2)^l  \, \bar{c}^{(p-l)}(\partial, \tdt+l).
			\end{aligned}
		\end{equation}
	Note that in the stationary case all coefficients $\bar{c}^{[p]}$, $\bar{d}^{[p]}$ are local. This is because for $k_0=0$ all ratios $k^2/\bk^2=1$ in the table of Appendix~\ref{sec:app_cfs_ans}, responsible for their spatial nonlocality, trivialize whereas the overall nonlocal factors of the form $1/k^2$ of $\bar{c}^{(p)}$ are compensated by suitable power of $k^2 \leftrightarrow {-}\partial^2$ in (\ref{eq:cfs_subtr}).
	
	Thus, the terms corresponding to the first sum in (\ref{eq:cfs_fun_form_stationary}), i.e.\ belonging to $\tau^2$-graded quadratic in $h_{\mu\nu}$ part of the heat kernel, are separately invariant (cf.\ discussion after (\ref{eq:cfs_fun_form})). In their turn, the sum of the terms of $\Tr K_2(\tau)$, corresponding to the $\tau^0$ and $\tau^1$ grading parts of (\ref{eq:cfs_fun_form_stationary}) supplemented by the terms of $\Tr K_1(\tau)$ of the same grading, are also invariant separately. Therefore, one should split the heat trace into the parts according to their grading and find for each of them the most general manifestly diffeomorphism invariant ansatz. In the stationary case the diffeomorphisms (\ref{eq:diffeo_def}) should be restricted to static ones with $\zeta^\mu = \zeta^\mu(\bx)$. This implies that $\bar{\xi}^\mu = \delta^\mu_0$ is itself invariant under such diffeomorphisms and is admissible for the construction of invariants.
	
	The $\tau^0$ and $\tau^1$ parts of the heat kernel, corresponding to the second sum in (\ref{eq:cfs_fun_form_stationary}), are local in the stationary case (i.e.\ have a polynomial dependence on $\partial^2$). Local $SO(d)$-invariant functionals associated with these terms and having zeroth and second order dimensionality in units of inverse length (complementary to the dimensionalities of $\tau^0$ and $\tau^1$) are composed of the dimensionless $g_{00}$, its derivatives $(\nabla_\mu g_{00})^2$, the potential $P$ and the first order in curvature quantities $R$ and $R_{00}$. Other scalar invariants built of spatial components of the curvature are expressible in terms of the latter two ones. Therefore, they can be taken in the form of $1$, in the $\tau^0$-grading part and
		\begin{equation} \label{eq:01_inv_bas}
			R,\quad R_{00},\quad (\nabla_\mu \omega)^{2},\quad P
		\end{equation}
	in the $\tau^1$ part of the total heat kernel trace. Here, for conformity with the notations of \cite{Dowker:1988jw} we denote by $\omega$ the logarithmic function of $g_{00}$,  $\omega = \frac12 \ln g_{00}$. Weighted by dimensionless local coefficients $u_I\equiv u_I(g_{00},\tau)$, $I=0,\ldots 4$, and integrated with the covariant Riemannian measure these invariants yield the ansatz for the covariant $\tau^0$ and $\tau^1$-graded part of the full heat trace, $\Tr K^{(0+1)}=\Tr K^{(0)}+\Tr K^{(1)}$,
		\begin{align}
				&\Tr K^{(0+1)}_{\rm covariant}[g] = \frac1{(4\pi \tau)^{D/2}}\int d^D x \sqrt{g} \,\Bigl\{ u_0			\nonumber\\&\hspace{1.75em}+\frac12\tau\Bigl[ u_1 R + u_2 R_{00} + u_3 \, (\nabla_\mu \omega)^{2} + u_4 P\, \Bigr]\,\Bigr\}. \label{eq:R1_inv}
		\end{align}
	Since $g_{00}$ is the only nontrivial dimensionless scalar of the stationary setup, the functional freedom in this ansatz consists in the above coefficient functions of $g_{00}$ and the proper time $\tau$.
	
	The covariant ansatz for the $\tau^2$-graded part, corresponding to the first sum of form factors in (\ref{eq:cfs_fun_form_stationary}), relies on invariants built of quadratic combination of curvatures and the potential $P$, because it starts with the quadratic order in $h_{\mu\nu}$ and the flat-space expansion of every curvature tensor begins with the linear order in metric perturbation. Moreover, each metric curvature is by itself diffeomorphism invariant under (\ref{eq:diffeo_def}) up to higher perturbation theory orders, $\delta R_{\mu\nu\rho\sigma} = \mathcal O(h^2)$. The complete basis of the curvature squared $SO(d)$-invariants (the potential $P$ being included into the set of ``curvatures'') can be chosen as
		\begin{equation} \label{eq:quad_inv_bas}
			R_{\mu\nu}^2, \;\; R^2, \;\; R P, \;\; P^2, \;\; R\hspace{0.1em}\bR, \;\; P\bR, \;\; \bR_{ij}^2, \;\; \bR^2,
		\end{equation}
	where $\bR$, $\bR_{ij}$ are $d$-dimensional Ricci scalar and tensor, respectively, constructed using spatial part of the metric. As with the time tensor components, the $d$-dimensional quantities here can be expressed via the $D$-dimensional Riemann, $R_{\alpha\beta\gamma\delta}\equiv {}^{(D)\!}R_{\alpha\beta\gamma\delta}$, and Ricci $R_{\mu\nu}\equiv{}^{(D)\!}R_{\mu\nu}$ tensors with the aid of the projector
	\begin{equation}
	    {\mathcal{P}}^\mu_\nu = \delta^\mu_\nu - \frac{{\xi}^\mu {\xi}_\nu}{\xi^2} \label{barprojector}
	\end{equation}
	taken at $\xi^\mu=\bar\xi^\mu$. In view of the Gauss-Codazzi equation $\bR = R - 2 R_{00}+\mathcal{O}[h^2]$ and $\bR_{ij} = R_{ij} -  R_{0i0j}+\mathcal{O}[h^2]$ the last four invariants in (\ref{eq:quad_inv_bas}) can be rewritten as bilinear combinations of $P$ and
	\begin{equation}
		\begin{aligned}
    		&\bR=R-2R_{\mu\nu}\frac{\xi^\mu\xi^\nu}{\xi^2}+\mathcal{O}[h^2], \\
    		&\bR_{\mu\nu}=R_{\alpha\beta}{\mathcal{P}}^\alpha_\mu {\mathcal{P}}^\beta_\nu - R_{\alpha\beta\gamma\delta}{\mathcal{P}}^\alpha_\mu {\mathcal{P}}^\gamma_\nu \frac{\xi^\beta\xi^\delta}{\xi^2}+{\cal O}[h^2],
		\end{aligned}
	\end{equation}
	also evaluated at $\bar\xi^\mu$. In particular, $\bR_{ij}^2=\bR_{\mu\nu}^2+{\cal O}[h^3]\equiv g^{\mu\alpha}g^{\nu\beta}\bR_{\mu\nu}\bR_{\alpha\beta}$.
	
	Completeness of the suggested basis (\ref{eq:quad_inv_bas}) follows from the observation that its expansion in $h_{\mu\nu}$, starting with a quadratic order, should saturate all admissible $SO(d)$-invariant combinations bilinear in metric perturbations and the potential $P$. They are determined by the set of $SO(d)$-invariant tensors $(\varPi_a)_{\mu\nu,\rho\sigma}$ and four basis tensors $(\varPi_b)_{\mu\nu}$ in the quadratic order of the noncovariant expansion (\ref{eq:TrK2}), listed in Appendix~\ref{sec:app_basis_tens} and restricted by the Ward identities (\ref{eq:Ward_ids}). For a nonlocal part these identities reduce to transversality conditions $k^\sigma \varPi_{\mu\nu,\rho\sigma} = 0$, $k^\sigma \varPi_{\rho\sigma} = 0$ and give rise to five independent structures for $\varPi_{\mu\nu,\rho\sigma}$ and two structures for $\varPi_{\rho\sigma}$. When supplemented with the potential square $P^2$ this gives a total of eight quadratic invariants.
		
	The $\tau^2$-graded part in question is essentially nonlocal, so that the coefficients of the quadratic invariants in $\Tr K^{(2)}$ are nonlocal form factors $f_I(\nabla)$, $I=1,\ldots 8$---nonlinear functions of covariant derivatives acting on one of the curvatures in their bilinear combinations,
		\begin{equation} \label{eq:R2_inv}
			\begin{aligned}
				&\Tr K^{(2)}_{\rm covariant}[g] = \frac{\tau^2}{(4\pi \tau)^{D/2}}\int d^Dx \sqrt{g} \Bigl[
				R_{\mu\nu} \, f_1(\nabla) \, R_{\mu\nu}\\
&\hspace{2em}+ R \, f_2(\nabla) \, R + P \, f_3(\nabla) \, R + P \, f_4 \, P+ R \, f_5(\nabla) \, \bR   \\
	    &\hspace{2em}+ P \, f_6(\nabla) \, \bR
	    + \bR^{\mu\nu} f_7(\nabla) \, \bR_{\mu\nu} + \bR \, f_8(\nabla) \, \bR\,
				\Bigr]
			\end{aligned}
		\end{equation}
	
	As discussed above, the $\tau^0$, $\tau^1$ and $\tau^2$-graded parts of the manifestly noncovariant expansion (\ref{eq:TrK1}),~(\ref{eq:TrK2}) are separately diffeomorphism invariant. Therefore, separately equating them to the manifestly covariant expressions (\ref{eq:R1_inv}) and (\ref{eq:R2_inv}) expanded to the second order in powers of $h_{\mu\nu}$, one obtains the system of equations for the coefficient functions $u_I(g_{00},\tau)$, $I=0,\ldots 4$, and form factors $f_I(\nabla)$, $I=1,\ldots 8$. In the overall quadratic order approximation in $h_{\mu\nu}$ these equations apply to second order in metric perturbations for the function $u_0(g_{00},\tau)$, to the first order for $u_I(g_{00},\tau)$, $I=1,\ldots 4$, and to zeroth order in $h_{\mu\nu}$ for $f_I(\nabla)$, $I=1,\ldots 8$.
	
	For the $\tau^0$, $\tau^1$-graded equations their solution can be written down, modulo some inessential freedom in $u_2$ and $u_3$, as
	    \begin{equation} \label{eq:u14_cfs}
	        \begin{aligned}
	        &u_0= \bigl[1 - h_{00} \, \tdt + \frac12 h_{00}^2 \, \tdt (\tdt+1)\,\bigr]\,r(\tau)+\mathcal{O}[h^3],\\
			&u_1= \frac16 (1- h_{00}\tdt)\,r(\tau)+\mathcal{O}[h^2], \\
			&u_2 = \frac13 \bigl(1 - h_{00}(\tdt+1)\bigr)  \,\tdt\,r(\tau)+\mathcal{O}[h^2],\\
			&u_3= \frac13 \tdt (\tdt+1)\,r(\tau)+\mathcal{O}[h^2],\\
			&u_4 = -(1 - h_{00}\tdt)\,r(\tau)+\mathcal{O}[h^2].
	       \end{aligned}
		\end{equation}
	The freedom in the choice of solutions for $u_2$ and $u_3$ is resolved by comparison with the non-perturbative (in metric perturbation) local Schwinger-DeWitt expansion for the thermal effective action achieved in \cite{Dowker:1988jw}. The same comparison also suggests a nonperturbative in $h_{\mu\nu}$ completion of the above equations
		\begin{align} \label{eq:I01_us}
				&u_0 = \frac{1}{(g_{00})^{\tdt}} r(\tau), \nonumber\\
				&u_1 = \frac16 \frac{1}{(g_{00})^{\tdt}}  r(\tau), \;\;
				u_2 = \frac{1}{3}\tdt\frac{1}{(g_{00})^{\tdt+1}}r(\tau), \\
				&u_3 = \frac{1}{3}  \frac{1}{(g_{00})^{\tdt}}\tdt(\tdt+1) r(\tau),  \;\;
				u_4 = - \frac{1}{(g_{00})^{\tdt}}  r(\tau). \nonumber
		\end{align}
		
	The same procedure applied to the manifestly covariant ansatz (\ref{eq:R2_inv}) and the relevant $\tau^2$-grading part of the expansion  (\ref{eq:TrK2}) (with the known form factors (\ref{eq:Pi_ansatz}) and their subtracted counterparts (\ref{eq:cfs_fun_form_stationary})) leads in the stationary case to the explicit functions $f_I=f_I(\tau,\partial)$. All these functions are expressed via the subtracted versions of the basic form factor function $f^{[p]}(\tau,\partial)$ with $p=0,1,2$. For the $SO(D)$-invariant curvature structures they turn out to be equal to
		\begin{equation} \label{eq:FF_stationary}
			\begin{aligned}
			&f_1 = -4(\tdt+2) f^{[2]} - f^{[1]},\\
			&f_2 = (\tdt+2)(\tdt+3) f^{[2]} + \frac12 f^{[1]},\\
			&f_3 = 2(\tdt + 1) f^{[1]},	\\
			&f_4 = f, \vphantom{\frac12}
		\end{aligned}
		\end{equation}
	while for the $SO(d)$-invariant ones they read
		\begin{equation} \label{eq:FF_stationary1}
			\begin{aligned}
			&f_5 = -(\tdt+2)(2\tdt+5) f^{[2]}  -\frac12 (\tdt+2) f^{[1]}, \hspace{-1em}\\
			&f_6 = -(2\tdt+3) f^{[1]}-\frac12 f,\\
			&f_7 =2 (2\tdt+5) f^{[2]} + f^{[1]} \phantom{\frac12}\\
			&\begin{aligned}
				&f_8 = \frac14(2\tdt+3)(2\tdt+5) f^{[2]} \\ &
				\hspace{5em}+ \frac14 (2\tdt+3) f^{[1]} + \frac1{16}f.
			\end{aligned}
		\end{aligned}
		\end{equation}
		
Thus, the final answer for the {\em stationary} heat kernel trace reads
	\begin{widetext}
		\begin{align}\label{eq:TrK_stationary}
				\Tr\bar{K}(\tau) &{}= \frac1{(4\pi\tau)^{D/2}} \int d^D x \sqrt{g} \biggl[ \frac{1}{(g_{00})^{\tdt}} r(\tau)
+\tau \frac{1}{(g_{00})^{\tdt}}\biggl(\frac16 R - P + \frac{1}{3}\frac{R_{00}}{g_{00}}\,\tdt
	    + \frac{1}{3} (\nabla\omega)^2\tdt(\tdt+1)\biggr) r(\tau) \nonumber\\
&\hspace{10em}+\frac12\tau^2\Bigl(R^{\mu\nu} \, \bar{f}_1(\nabla)R_{\mu\nu} + R \, \bar{f}_2(\nabla)R
	    + P \, \bar{f}_3(\nabla)R + P \, \bar{f}_4(\nabla)P
\\ & \hspace{10em}
+ R \, \bar{f}_5(\nabla)\bR + P \, \bar{f}_6(\nabla)\bR + \bR^{\mu\nu} \, \bar{f}_7(\nabla)\bR_{\mu\nu}
	    + \bR \, \bar{f}_8(\nabla)\bR  \Bigr)\biggr]+{\cal O}[\Re^3], \nonumber
	\end{align}
	\end{widetext}
	where, of course, $\bar f_I(\nabla)=f_I(\tau,\nabla)|_{\partial_0=0}$ and the replacement of $\partial$ by the covariant derivative $\nabla$ along with the replacement of $\partial^2$ by the covariant d'Alembertian $\Box=g^{\mu\nu}\nabla_\mu\nabla_\nu$ (note that the fundamental form factor~(\ref{eq:f_def}) with $\partial_0=0$ depends only on $\partial^2$) can be freely done because relevant corrections exceed the assumed quadratic order precision in curvatures.

	The origin of the differential in $\tau$ operator $(g_{00})^{-\tdt}$, which is supposed to be acting to the right, is closely related here to the notion of Tolman temperature defined as $T_{\text{Tolman}} = T / \sqrt{g_{00}}$. Indeed, this operator acts on the function $r(\tau)$ by scaling its argument by a factor of $(g_{00})^{-1}$, i.e.\ $(g_{00})^{-\tdt}r(\tau) =  r(\tau / g_{00})$. According to the definition (\ref{eq:r_def}) of the function $r(\tau)$, its argument $\tau$ enters only in the combination $\tau / \beta^2 = T^2 \tau$. Consequently, this effectively maps the global temperature $T$ to the local Tolman one, namely
		\begin{equation}
			(g_{00})^{-\tdt} r(\tau) = r(\tau)\Bigr|_{T=T_{\text{Tolman}}},
		\end{equation}
	which fully confirms the result of \cite{Dowker:1988jw}.

\subsection{Generalized Killing vector and covariant form in the non-stationary case}
	Transition to a non-stationary case is associated with several new features of the formalism. First, the basic form factor (\ref{eq:f_def}) no longer factorizes like in (\ref{eq:ff_stationary}), so that its subtraction procedure, necessary for $\tau^p$-grading of various terms, should be modified as
		\begin{equation} \label{eq:ff_subtr}
			f^{[p]}(\tau, \partial) = \frac1{(-\tau \partial^2)^p}\Bigl(f(\tau, \partial)
	    -  \sum_{l=0}^{p-1}\frac{l!}{(2l+1)!} r(\tau) (\tau \partial^2)^l  \Bigr).
		\end{equation}
	This again allows one to rewrite the coefficients (\ref{eq:cfs_fun_form}) of the ansatz (\ref{eq:Pi_ansatz}) in the form
		\begin{align} \label{eq:cfs_fun_form_subtr}
			C(\tau, \partial) = \tau^2 \sum_{p=0}^2 c^{[p]}(\partial,\tdt) f^{[p]}(\tau,\partial)
	    \nonumber\\[-1ex]
        + \sum_{p=0}^1 \tau^p \, d^{[p]}(\partial,\tdt)\, r(\tau),
		\end{align}
	where the coefficients $c^{[p]}$,~$d^{[p]}$ are expressed in terms of $c^{(p)}$,~$d^{(p)}$ by the same relations (\ref{eq:cfs_subtr}) as in the stationary case. However, now all these coefficients are essentially nonlocal because of the factors of $k^2/\bk^2=1+k_0^2/\bk^2$ in the table of their momentum representation expressions of Appendix~\ref{sec:app_cfs_ans}.
	
	The second most important feature of the non-stationary formalism is that $\bar{\xi}^\mu = \delta^\mu_0$ is no longer a fixed vector of Killing isometry, and if we want to arrange diffeomorphism invariant quantities by replacing temporal indices in terms of tensor contractions with some vector field $\xi^\mu$, this vector should  have a correct diffeomorphism transformation property under the transformation (\ref{eq:diffeo_def}) of $h_{\mu\nu}$. Therefore, this vector field is necessarily some functional of $h_{\mu\nu}(x)$, $\xi^\mu=\xi^\mu[h]$, which reduces to the Killing field $\bar{\xi}^\mu$ on a stationary background. In a special coordinate system in which the metric perturbation $h_{\mu\nu}(x)=h_{\mu\nu}(\bx)$ is time independent, this means that it should satisfy the condition $\xi^\mu[h(\bx)]=\delta^\mu_0$.
	
	We will look for such a generalization of the Killing vector field in the form of the $h_{\mu\nu}$-expansion
	\begin{equation}\label{eq:xi}
		\begin{aligned}
			&\xi^\mu[h] = \delta_0^\mu + \varXi^\mu_{\nu\rho}(\partial) h^{\nu\rho}(x) \\ &\;+ \frac12\varXi^\mu_{\nu\rho,\sigma\kappa}(\partial_x,\partial_y) h^{\nu\rho}(x) h^{\sigma\kappa}(y)\bigr|_{y=x}+\mathcal{O}[h^3],
		\end{aligned}
	\end{equation}
	with some unknown operator coefficients $\varXi^\mu_{\nu\rho}(\partial)$ and $\varXi^\mu_{\nu\rho,\sigma\kappa}(\partial_x,\partial_y)$. These coefficients are restricted by the requirement of the vector-field diffeomorphism of $\xi^\mu[h]$ under the diffeomorphism (\ref{eq:diffeo_def}) of $h_{\mu\nu}$. In the momentum space representation this requirement implies the Ward type identities
	\begin{align}
		&k^\nu\varXi^\mu_{\nu\rho}(ik) = -\frac12 k_0  \delta^\mu_\rho, \\
		&\begin{aligned}
			&k^\nu \varXi^\mu_{\nu\rho,\alpha\beta}(ik,i(p-k)) = - \frac12 \Bigl[ (p-k)_\rho \varXi^\mu_{\alpha\beta}(ip) \\&\hspace{5.5em}+ \delta_{\beta \rho} k^\nu \varXi^\mu_{\nu\beta}(ip) + \delta_{\alpha \rho}k^\nu \varXi^\mu_{\alpha\nu}(i p)  \Bigr] \\
			&\hspace{1em}+ \frac12 (p-k)_\rho \varXi^\mu_{\alpha\beta}(i(p-k)) - \frac12 \delta^\mu_\rho k_\nu \varXi^\nu_{\alpha\beta}(i(p-k))\hspace{-2em}
		\end{aligned}
	\end{align}
	which, of course, do not fix these coefficients completely. The above requirement of a well-defined limit to the stationary case, $\xi^\mu[h(\bx)] = \bar{\xi}^\mu$, adds the condition that both of the operator coefficients should vanish at $\partial_0=0$, $\varXi|_{\partial_0=0} = 0$.
	This condition also ensures that the vector field $\xi^\mu$ generates periodic integral curves, $dx^\mu(x^0)/dx^0=\xi^\mu(x(x^0))$, $\int_0^\beta dx^0\,\xi^\mu(x(x^0))=\beta\delta^\mu_0$, along which periodicity of the underlying manifold takes place.
	The rest of the information on the choice of $\varXi^\mu_{\nu\rho}(\partial)$ and $\varXi^\mu_{\nu\rho,\sigma\kappa}(\partial_x,\partial_y)$ should follow from the non-stationary version of covariantization procedure, in which the determination of $\xi^\mu[h]$ gets on equal footing with the above determination of the coefficient functions of the ansatzes (\ref{eq:R1_inv}) and (\ref{eq:R2_inv}).
	
	The covariant ansatzes (\ref{eq:R1_inv}) and (\ref{eq:R2_inv}) should also undergo a modification. Apart from obvious replacements of temporal indices by contractions with the introduced vector field, that is $g_{00} \mapsto g_{\xi\xi} = g_{\mu\nu} \xi^\mu \xi^\nu$, $R_{00} \mapsto R_{\xi\xi} = R_{\mu\nu} \xi^\mu \xi^\nu$, additional tensor invariants turn out to be necessary. This follows from the $\tau^0$ and $\tau^1$ graded parts of the total answer, which as discussed above are separately diffeomorphism invariant. One might have expected that these parts could be expressed in terms of the covariant ansatz (\ref{eq:R1_inv}), which, in particular, would imply that all nonlocality and breakdown of Lorentz invariance due to the presence of $\bar\xi^\mu=\delta^\mu_0$, could be hidden inside the definition of the vector (\ref{eq:xi}) under an appropriate choice of its coefficients $\varXi$. However, this expectation turns out to be incorrect.
	
	Indeed, expanding in $h_{\mu\nu}$ the covariantized by the vector $\xi^\mu$ invariant (\ref{eq:R1_inv}) with the coefficient functions (\ref{eq:I01_us}), which we assume to be the same as in the stationary case, and equating it to the $\tau^0$ and $\tau^1$ graded part of the original noncovariant expansion, one obtains the system of equations for $\varXi$. However, this system turns out to be overdetermined.	To circumvent this difficulty one can use the freedom to include into (\ref{eq:R1_inv}) additional invariant quantities of second order in derivatives,
		\begin{equation} \label{eq:Xi_inv}
			\begin{aligned}
				&\bigl(\nabla_{(\mu} \xi_{\nu)}\bigr)^2, \qquad &&\bigl(\nabla_\mu\xi^\mu\bigr)^2, \\
	            &\xi^\mu\xi^\alpha\nabla_\alpha\xi^\nu \nabla_{(\mu}\xi_{\nu)}, \qquad
				&&\xi^\mu\nabla^\nu \xi^2 \, \nabla_{(\mu}\xi_{\nu)}, \\
				&\xi^\mu \xi^\nu \nabla_{(\mu} \xi_{\nu)}\nabla_\rho\xi^\rho ,\qquad
				&&\bigl(\xi^\mu \xi^\nu \nabla_{(\mu} \xi_{\nu)}\bigr)^2,
			\end{aligned}
		\end{equation}
	which are all proportional to the left hand side of the Killing equation $\nabla{}_{(\mu} \xi{}_{\nu)}$. These invariants obviously vanish in the stationary case, because $\xi^\mu$ becomes a Killing vector $\bar{\xi}^\mu$ for the stationary perturbations. Inclusion of these combinations with free coefficients leads to a consistent set of equations on $\varXi^\mu_{\nu\rho}(\partial)$ and $\varXi^0_{\mu\nu,\rho\sigma}(\partial,{-}\partial)$. Note that spatial components of $\varXi^\mu_{\nu\rho,\sigma\kappa}$ unlike the temporal one $\varXi^0_{\mu\nu,\rho\sigma}(\partial,{-}\partial)$ cannot be determined at the quadratic order of perturbation theory, because it arises only in the $u_0(g_{\xi\xi})$-term of~(\ref{eq:R1_inv}), specifically in the integrated form of its temporal component, $\int d^Dx \, h_{\mu\nu}(x) \varXi^0_{\mu\nu,\rho\sigma}(\partial, -\partial) h_{\rho\sigma}(x)$.
		
	Thus, by adding the invariants (\ref{eq:Xi_inv}) weighted by unknown coefficients into the integrand of (\ref{eq:R1_inv}) one has a new system of equations. It governs not only the coefficient functions $\varXi$, but also these coefficients which define the weights of the new terms in the modified ansatz (\ref{eq:R1_inv}).	The resulting system turns out to be underdetermined and contains three free dimensionless parameters. In particular, the solution for $\varXi^\mu_{\nu\rho}(\partial)$ turns out to have the form
	\vspace{-1ex}
	\begin{align}
			&\varXi^\mu_{\nu\rho}(\partial) = \frac{\partial_0}{2\bpar^2 \partial^2}\biggl[ \partial^\mu (\partial_{\nu} \partial_{\rho} {-} \partial^2 \bar{\xi}_\nu \bar\xi_\rho) {-}  (\delta^\mu_\nu \bpar_\rho {+} \delta^\mu_\rho \bpar_\nu) \partial^2 \nonumber\\&
           	- a \frac{\bpar^\mu}{\bpar^2} (\partial_0 \partial_\nu {-}  \bar{\xi}_\nu\partial^2)(\partial_0 \partial_\rho {-}  \bar{\xi}_\rho\partial^2)
    		+b \,\bpar^\mu (\eta_{\nu\rho} \partial^2 {-} \partial_\nu \partial_\rho)
  		  \biggr],
	\end{align}	\vspace{-2ex}\\
	where $\bar{\mathcal{P}}^{\mu\nu} = \eta^{\mu\nu} - \bar{\xi}^\mu \bar{\xi}^\nu$ is the projector, and $\bpar^\mu = \bar{\mathcal{P}}^{\mu\nu} \partial_\nu$ is the spatial part of the partial derivative. Here $a$ and $b$ are two out of three free parameters of the solution. The solution for the coefficient function $\varXi^0_{\nu\rho,\sigma\kappa}(\partial,{-}\partial)$ is cumbersome and is given in Appendix~\ref{sec:app_Xi4}. The third free parameter enters the weights of the scalars (\ref{eq:Xi_inv}) which are dependent at the quadratic order, namely, its particular combination vanishes
	\begin{align}
		&(b+1)\,\xi^\mu \, \nabla^\nu \xi^2 \, \nabla_{(\mu}\xi_{\nu)} -
		b\,\xi^\mu \xi^\nu \nabla_{(\mu} \xi_{\nu)} \, \nabla_\rho\xi^\rho \nonumber\\ &\hspace{4em}+
		(a-b) \, \bigl(\xi^\mu \xi^\nu \nabla_{(\mu} \xi_{\nu)}\bigr)^2 = {\cal O}[h^3]. \label{identity}
	\end{align}
	Hence, this combination can be added to the particular form of the final answer with an arbitrary coefficient, which is the third parameter of the solution. The ambiguity in the choice of this dimensionless coefficient is of functional nature---it can be the function of the operator $\tdt$ acting on $r(\tau)$. This is in contrast to numerical coefficients $a$ and $b$, which cannot involve $\tdt$ because the derivation procedure for the equations on $\varXi$ and the weights of (\ref{eq:Xi_inv}) was strongly facilitated by the additional grading of all terms in powers of the operator $\tdt$.
	
	Covariantization of the $\tau^2$-graded part of the heat kernel in terms of the ansatz (\ref{eq:R2_inv}) does not require a replacement of $\bar\xi^\mu$ by $\xi^\mu$ when working at the quadratic order of metric perturbations since the difference between $\xi^\mu$ and $\bar{\xi}^\mu$ affects only higher order terms. Therefore, it basically repeats the procedure of the stationary case---rewriting (\ref{eq:R2_inv}) in terms of quadratic combinations of $h_{\mu\nu}$ and equating it to $\tau^2$-graded part of the expansion  (\ref{eq:TrK2}) (with the known form factors (\ref{eq:Pi_ansatz}) and their subtracted {\em non-stationary} counterparts (\ref{eq:cfs_fun_form_subtr}). Just like the stationary case this leads to the explicit non-stationary functions, which we temporarily denote by $f_I^{\rm nonstat}(\tau,\partial)$, expressed via the subtracted versions of the basic non-stationary form factor function $f^{[p]}(\tau,\partial)$ given by (\ref{eq:ff_subtr}), $p=0,1,2$. For the $SO(D)$-invariant curvature structures with $I=1,\ldots 4$ these expressions in terms of $f^{[p]}$ exactly coincide with those of Eq.~(\ref{eq:FF_stationary}), while for the $SO(d)$-invariant structures, $I=5,\ldots 8$, these expressions are given by Eqs.~(\ref{eq:FF_stationary1}) modified by extra nonlocal factors $k^2/\bk^2=1+k_0^2/\bk^2$ originating in the non-stationary case from the table of Appendix~\ref{sec:app_cfs_ans}. In the momentum space representation the above relations read
		\begin{align}
			&f_I^{\rm nonstat}(\tau,ik) = f_I(\tau,ik), && I=1,\ldots 4,  \label{k1}\\
			&f_I^{\rm nonstat}(\tau,ik) = (k^2/\bk^2) f_I(\tau,ik), && I=5,6, \label{k2}\\
			&f_I^{\rm nonstat}(\tau,ik) = (k^2/\bk^2)^2 f_I(\tau,ik), && I=7,8.  \label{k3}
		\end{align}
	Collecting all the above results together, one finds that the final answer for non-stationary heat kernel trace can be rendered in a particular form
	\begin{widetext}
		\begin{equation} \label{eq:TrK_cov}
			\begin{aligned}
				\Tr \,&K(\tau) =  \frac1{(4\pi\tau)^{D/2}}\int d^D x \sqrt{g} \biggl\{ \frac{1}{(g_{\xi\xi})^{\tdt}} r(\tau)
	            +\tau \frac{1}{(g_{\xi\xi})^{\tdt}}\biggl[\frac16 R - P \\
	            &\hspace{1em}+ \frac{1}{3} \biggl(\frac{R_{\xi\xi}}{g_{\xi\xi}} + \bigl(\nabla_{(\mu}\xi_{\nu)}\bigr)^2
	            -  \bigl(\nabla_\mu \xi^\mu\bigr)^2 - \xi^\mu \, (\nabla_\xi \xi^\nu) \nabla_{(\mu}\xi_{\nu)} + \frac32 \xi^\mu \, (\nabla^\nu \xi^2) \, \nabla_{(\mu}\xi_{\nu)}\biggr)\tdt\\
	            &\hspace{1em}+ \frac{1}{3} \Bigl( (\nabla_\mu\omega)^2 -2 \xi^\mu \, (\nabla^\nu \xi^2) \, \nabla_{(\mu}\xi_{\nu)} + 2\xi^\mu \xi^\nu (\nabla_{(\mu}\xi_{\nu)}) \, \nabla_\rho\xi^\rho \Bigr) \tdt(\tdt+1) \biggr] r(\tau)  \\
	            &\hspace{1em}+
				\frac12\tau^2\Bigl(R_{\mu\nu}f_1(\nabla)R_{\mu\nu} + R f_2(\nabla)R + P f_3(\nabla)R + P f_4(\nabla)P \\
	            & \hspace{3em}+ R\, f_5(\nabla) \mfrac{\Box}{\Delta} \bR  + P\, f_6(\nabla)\mfrac{\Box}{\Delta}\bR
	            + \bR^{\mu\nu}\,\mfrac{\Box}{\Delta} f_7(\nabla)\mfrac{\Box}{\Delta} \bR_{\mu\nu}
	            + \bR\mfrac{\Box}{\Delta} f_8(\nabla)\mfrac{\Box}{\Delta} \bR \Bigr)\biggr\}+\mathcal{O}[\Re^3].
			\end{aligned}\hspace{-2.3em}
		\end{equation}
	\end{widetext}
	Here new structures (\ref{eq:Xi_inv}) with their explicit coefficients are contained in big parentheses of the second and third lines. Covariantly completed form factors of the quadratic invariants follow from the momentum space functions (\ref{k1})--(\ref{k3}) in terms of the
	set of functions $f_I(\tau,\partial)$ of Eqs.~(\ref{eq:FF_stationary}),~(\ref{eq:FF_stationary1}),
	\begin{equation}
		f_I(\nabla)=f_I(\tau,\partial)\bigr|_{\partial\,\mapsto\nabla}
	\end{equation}
	where we omit explicit $\tau$-dependence for the sake of conciseness.
	They include $D$-dimensional d'Alembertian~$\Box$ and $d$-dimensional Laplacian, defined as
	\begin{equation}
		\Delta=\nabla_\mu\mathcal{P}^{\mu\nu}\nabla_\nu,
	\end{equation}
	where $\mathcal{P}^{\mu\nu}$ is the non-stationary version of the projector (\ref{barprojector}) built in terms of the generalized Killing vector $\xi^\mu$. Covariant completion with noncommuting objects, $k_\mu\mapsto-i\nabla_\mu$, $k^2\mapsto-\Box$, $\bk^2\mapsto-\Delta$ and $\mathcal{P}^{\mu\nu}$ is organized in such a way that the spatial Laplacian $\Delta$ as well as the kernels of quadratic curvature forms represent operators symmetric under spacetime integration with the Rimannian measure---this explains symmetric ordering in the definition $\Delta$ and in the last quadratic terms of (\ref{eq:TrK_cov}). These details might be inessential in the quadratic order approximation, but affect higher orders of covariant perturbation theory, if those orders will be chosen to proceed with---like this happens with the cubic curvature order in the vacuum case \cite{Barvinsky:1993en}.
	
	As mentioned above, there is a three parameter freedom in the representation of (\ref{eq:TrK_cov})---two parameters $a$ and $b$ in the construction of $\xi^\mu[h]$ and extra parameter associated with the identity (\ref{identity}). From the above derivation procedure it is obvious that the terms inside big parentheses of the second and third lines of (\ref{eq:TrK_cov}) are independent of $a$ and $b$, parameterizing the generalized Killing vector, while the coefficients of the interior $\xi^\mu$-structures can be altered by the addition of this identity with the third coefficient which can be a generic function of the differential operator $\tdt$ acting on $r(\tau)$. This leads to another form of the final result explicitly featuring all three arbitrary parameters, but we prefer not to do that in (\ref{eq:TrK_cov}).
		
	A final comment to Eq.~(\ref{eq:TrK_cov}) is that its notations are somewhat redundant, namely $g_{\xi\xi} \equiv \xi^2 \equiv e^{2 \omega}$. Moreover, some of the terms derived from the Killing equation can also be rewritten in terms of $\omega$, for example, $\xi^\mu \nabla_\nu \xi_\mu = \frac{1}{2} \nabla_\nu \xi^2 = \nabla_\nu \omega + \mathcal{O}(h^2)$. Nevertheless, the final answer is retained in this form to maintain key aspects, namely a closer analogy to the static case (\ref{eq:TrK_stationary}) and thermal results of Dowker and Schofield \cite{Dowker:1988jw}.

\subsection{Separation of the vacuum contribution} \label{sec:vac_separ}
	It turns out that the answer for the heat kernel trace can be effectively separated into the vacuum part, corresponding to the zero temperature (or infinite $\beta$), and the pure thermal part. To show this, we use the Poisson summation formula to the sum in the definition (\ref{eq:f_def}) of the form factor $f(\tau, \partial)$, that gives
	\begin{equation} \label{eq:poisson}
		\frac{\sqrt{4\pi\tau}}{\beta}\sum_{n} e^{-\tau (\omega_n+i\alpha\partial_0)^2}
        = 1+2\sum_{n=1}^\infty e^{-\frac{\beta^2 n^2}{4\tau}}\cosh(n\alpha \beta \partial_0).
	\end{equation}
	Hence, the thermal function and the fundamental form factor split into the parts, respectively, referred to as the vacuum and thermal ones,
	\begin{align}
        &r(\tau) = 1 + r_\beta(\tau), \qquad r_\beta(\tau) = 2\sum_{n=1}^\infty e^{-\frac{\beta^2n^2}{4\tau}}, \label{eq:r_split}\\
		&f(\tau, \partial) = f_{\mathrm{vac}}(-\tau \partial^2) + f_\beta(\tau, \partial), \label{eq:f_split}\\
        &f_\beta(\tau, \partial) = 2\int_0^1 d\alpha \,e^{\alpha(1-\alpha) \tau \partial^2}
        \sum_{n=1}^\infty e^{-\frac{\beta^2 n^2}{4\tau}} \cosh(n\alpha \beta \partial_0), \label{eq:f_beta}
	\end{align}
	and corresponding to the first and second terms in the right hand side of (\ref{eq:poisson}). The explicit form of the vacuum form factor $f_{\mathrm{vac}}(-\tau \partial^2)$ was previously defined in (\ref{vac_ff}), whereas the $f_\beta(\tau, \partial)$ represents the fundamental thermal form factor. The vacuum form factor is the function of the combination $-\tau \partial^2$, whereas the thermal one is the function of the separate variables $\tau$ and $\partial_\mu = (\partial_0, \bpar)$.
		
	This splitting induces the corresponding decomposition of the heat trace into vacuum and thermal components based on the respective parts of the thermal function and the fundamental form factor,
		\begin{equation} \label{vac-therm}
	    \Tr K(\tau) = \Tr K_{\mathrm{vac}}(\tau) + \Tr K_\beta(\tau).
		\end{equation}
	
	Significant simplification takes place in the vacuum part $\Tr K_{\mathrm{vac}}(\tau)$. In its $\tau^0$ and $\tau^1$ graded terms---the first three lines of (\ref{eq:TrK_cov})---the operator $\tdt$ acts on $r_{\mathrm{vac}}(\tau)=1$, and these terms reduce to the local contributions of the first two Schwinger-DeWitt coefficients of the vacuum heat kernel (\ref{eq:intro_vacuum}), which are of zeroth and first order in Ricci curvature.
		
	The $\tau^2$-graded terms corresponding to the curvature-squared part of the heat kernel (\ref{eq:TrK_cov}) contain the form factors $f_I(\tau,\partial)$ which are expressed via Eqs.~(\ref{eq:FF_stationary}),~(\ref{eq:FF_stationary1}) in terms of the subtracted functions $f^{[p]}(\tau,\partial)$. These functions in turn get split into vacuum and thermal parts $f^{[p]}(\tau,\partial) = f^{[p]}_{\mathrm{vac}}(-\tau\partial^2) + f^{[p]}_\beta(\tau,\partial)$, which results in the vacuum-thermal decomposition of all form factors
		\begin{equation} \label{vacuum-thermal-f}
	    f_I(\tau,\partial)=f^{\mathrm{vac}}_I(\tau,\partial)+f^\beta_I(\tau,\partial), \quad I=1,\ldots 8.
	     \end{equation}
	The vacuum ones can be further simplified. As follows directly from the definition (\ref{vac_ff}), $f_{\mathrm{vac}}(u)$ satisfies the first order differential equation
		\begin{equation} \label{eq:f_vac_rec_1}
			u\partial_u f_{\mathrm{vac}}(u) = -\frac14 (u + 2) f_{\mathrm{vac}}(u)+ \frac12.
		\end{equation}
	This equation allows one to express the $\tdt$-derivatives of $f_{\mathrm{vac}}(-\tau\partial^2)$ solely in terms of $f_{\mathrm{vac}}(-\tau\partial^2)$. Upon the use of (\ref{eq:f_vac_rec_1}) the vacuum parts of the first four form factor functions $f^{\mathrm{vac}}_I(u)$, $I=1,\ldots 4$, reduce to
		\begin{equation} \label{eq:FF_vac}
	    \begin{aligned}
			&f^{\mathrm{vac}}_1(u) = 2 f_{\mathrm{vac}}^{[2]}(u),\\
		    &f^{\mathrm{vac}}_2(u) = -\frac{1}{4} f_{\mathrm{vac}}^{[2]}(u) + \frac{1}{4} f_{\mathrm{vac}}^{[1]}(u) + \frac{1}{16}f_{\mathrm{vac}}(u),\\
			&f^{\mathrm{vac}}_3(u) = -f_{\mathrm{vac}}^{[1]}(u) - \frac{1}{2}f_{\mathrm{vac}}(u),\\
			&f^{\mathrm{vac}}_4(u) = f_{\mathrm{vac}}(u),
	    \end{aligned}
		\end{equation}
	where $f_{\mathrm{vac}}^{[p]}(u)$ are the subtracted functions (\ref{eq:ff_vac_subtr}), while those of the $SO(d)$-invariant terms identically vanish, $f_I^{\mathrm{vac}}(u)=0$, $I=5,\ldots 8$. Thus, with $u=-\tau\Box$ one arrives at the  answer for the covariantly completed vacuum part of the heat trace (\ref{eq:intro_vacuum}) known from \cite{Barvinsky:1990up} in a more general case of a generic fibre bundle curvature ${\cal R}_{\mu\nu}$.
	
	In contrast to the vacuum form factors $f^{\mathrm{vac}}_I(-\tau\Box)$, the thermal form factors do not admit any additional identities like (\ref{eq:f_vac_rec_1}), so they cannot be further simplified. However, simplifications arise when these factors are used in effective action calculations, as described below.
	
\section{Effective action} \label{sec:effective_action}
	
	The one-loop effective action (\ref{effective_action}) for the wave operator $F$ reads within zeta-function regularization as
		\begin{equation} \label{eq:eff_act}
			W = -\frac12 \frac{\partial}{\partial s} \biggl[\frac{\mu^{2s}}{\Gamma(s)}\int_0^\infty d\tau \, \tau^{s-1} \Tr K(\tau)\biggr]\biggr|_{s=0}, \
		\end{equation}
	where $\mu$ is the normalization parameter reflecting renormalization ambiguity (and absorbing logarithmic UV divergences arising in dimensional regularization). Similarly to the heat kernel, effective action consists of the vacuum and thermal parts.
	
	In the quadratic order the vacuum part $W_{\mathrm{vac}}$ is generated only by $\tau^2$-graded terms of the heat kernel, because its $\tau^0$ and $\tau^1$ parts have purely power-law dependence on $\tau$ and get annihilated under zeta-function regularization, $\int_0^\infty d\tau \, \tau^q \equiv 0$ (as well as in the dimensional regularization). Thus, $W_{\mathrm{vac}}$ begins with the quadratic terms in curvatures, whose form factors $\phi^{\mathrm{vac}}_I(-\Box)$ follow from those of the $\tau^2$-part of $\Tr K_{\mathrm{vac}}$, $f^{\mathrm{vac}}_I(-\tau\Box)$, by the replacement
	\begin{align} \label{eq:ea_ff_vac}
		&f^{\mathrm{vac}}_I(-\tau\Box)\mapsto\phi^{\mathrm{vac}}_I(-\Box) \nonumber\\ &\;= \frac{\partial}{\partial s} \biggl[\frac{\mu^{2s}}{\Gamma(s)}\int_0^\infty d\tau \,
	    \tau^{s-1} \frac1{\tau^{D/2}} \tau^2 f^{\mathrm{vac}}_I(-\tau \Box) \biggr] \biggr|_{s=0}.
	\end{align}
	
	In view of the explicit relations (\ref{eq:FF_vac}), expressing $f^{\mathrm{vac}}_I(-\tau \Box)$ in terms of the basic vacuum form factor $f_{\mathrm{vac}}(-\tau \Box)$ and its subtracted counterparts, $f_{\mathrm{vac}}^{[p]}(-\tau \Box)\sim f_{\mathrm{vac}}(-\tau \Box)/(-\tau\Box)^p+\ldots$, the set of the above form factors are expressed as linear combinations of operator functions of the d'Alembertian $(-\Box)^{-p} \phi_{\mathrm{vac}}^{(D/2-2+p)}(-\Box)$ having one and the same dimensionality $D-4$. Here $\phi_{\mathrm{vac}}^{(q)}(-\Box)$ is the following set of basic effective action form factors with $q=0,1,2$,
	\begin{align}
		&\phi_{\mathrm{vac}}^{(q)}(-\Box) = \frac{\partial}{\partial s} \biggl[\frac{\mu^{2s}}{\Gamma(s)}
	    \int_0^\infty d\tau \, \tau^{s-q-1}  f_{\mathrm{vac}}(-\tau \Box) \biggr] \biggr|_{s=0} \nonumber\\ &\quad=
		-\frac{q!}{(2q+1)!} \, \Box^{q} \, \Bigl[\log\Bigl(-\mfrac{\Box}{\mu^2}\Bigr) + H_{q} - 2 H_{2q+1}\Bigr], \label{eq:vphi_vac_ans_k}
	\end{align}
	where $H_l = \sum_{n=1}^l 1/n$ is the harmonic number and in the last equality we assume the case of even dimensional spacetime $D$.
	
	Thus, the vacuum part of effective action takes the form
	\begin{align}
			&-W_{\mathrm{vac}} = \frac1{2(4\pi)^{D/2}} \int d^D x \sqrt{g}\nonumber \\ &\quad\times \Bigl( R^{\mu\nu} \phi^{\mathrm{vac}}_1(-\Box) R_{\mu\nu}
		      + R \, \phi^{\mathrm{vac}}_2(-\Box) R
		    \\ &\qquad\qquad+ P \, \phi^{\mathrm{vac}}_3(-\Box) \, R + P \, \phi^{\mathrm{vac}}_4 (-\Box)P \Bigr)+\mathcal{O}[\Re^3] \nonumber
	\end{align}
	with the form factors fully matching the old result of \cite{Barvinsky:1990up} obtained in dimensional regularization,
	\begin{equation}
		\begin{aligned} \label{eq:vac_ff_ea}
			&\phi^{\mathrm{vac}}_1(-\Box) = \mfrac2{(-\Box)^2} \phi_{\mathrm{vac}}^{(D/2)}(-\Box),\\
			&\begin{aligned}
				\phi^{\mathrm{vac}}_2(-\Box) = -\mfrac{1}{4\,(-\Box)^2} &\phi_{\mathrm{vac}}^{(D/2)}(-\Box) + \mfrac{1}{4\,(-\Box)} \phi_{\mathrm{vac}}^{(D/2-1)}(-\Box) \\+ \mfrac{1}{16}&\phi_{\mathrm{vac}}^{(D/2-2)}(-\Box),
	        \end{aligned} \\
			&\phi^{\mathrm{vac}}_3(-\Box) = -\mfrac1{(-\Box)}\phi_{\mathrm{vac}}^{(D/2-1)}(-\Box) - \mfrac{1}{2}\phi_{\mathrm{vac}}^{(D/2-2)}(-\Box),\\
			&\phi^{\mathrm{vac}}_4(-\Box) = \phi_{\mathrm{vac}}^{(D/2-2)}(-\Box).
		\end{aligned}
	\end{equation}
	
	The thermal part of the effective action is finite, i.e.\ the integral over $\tau$ in (\ref{eq:eff_act}) is convergent in $s \to 0$ limit. Hence the zeta-regularization can be lifted, and the thermal part of the effective action reads
	\begin{equation} \label{eq:eff_act_thermal}
		W_{\beta} =  -\frac12 \int_0^\infty \frac{d\tau}{\tau} \Tr K_{\beta}(\tau).
	\end{equation}
	Its $\tau^0$ and $\tau^1$ graded terms are of the form
	\begin{align} \label{eq:ibp_r_beta}
		\int_0^\infty \frac{d\tau}{\tau} \frac{1}{\tau^{D/2}} \, \tau^{p} \,& c(\tdt) \, r_\beta(\tau)\nonumber
    \\ =  &\;c(\tfrac{D}2-p)\int_0^\infty \frac{d\tau}{\tau} \frac{1}{\tau^{D/2}} \, \tau^{p} r_\beta(\tau)
    \\ =  &\;c(\tfrac{D}2 - p) \, \rho^{(D/2-p)}  \nonumber
	\end{align}
	with $p=0,1$, $c(\tdt)$ being some function of $\tdt$, and thermal function $r_\beta(\tau)$ defined by (\ref{eq:r_split}). Here in the first equality we integrated by parts (no boundary terms arise since $r_\beta(\tau)$ vanishes at $\tau=0$ along with all its derivatives), while in the second equality we exchanged the order of summation in the expression for $r_\beta(\tau)$ and integration followed by the use of Riemann zeta function $\zeta(s) = \sum_{n=1}^\infty 1/n^{s}$ in the calculation of the set of numerical coefficients,
	\begin{align} \label{eq:rho_def}
		\rho^{(q)} &= \int_0^\infty \frac{d\tau}{\tau^{q+1}} r_\beta(\tau) =
		2\Bigl(\frac{2}{\beta}\Bigr)^{\!2q} \Gamma(q) \zeta(2q).
	\end{align}
	
	In contrast to these numerical coefficients, the integration of thermal $\tau^2$-terms of the heat kernel (\ref{eq:TrK_cov}) leads to the {\em thermal effective action form factors} which are nonlocal operators
		\begin{equation} \label{eq:ea_ff_I}
			\phi^\beta_I(\partial) = \int_0^\infty \frac{d\tau}{\tau} \frac1{\tau^{D/2}} \tau^2 f^\beta_I(\tau, \partial)
		\end{equation}
	where thermal $f^\beta_I(\tau, \partial)$ are expressed via (\ref{eq:FF_stationary}),~(\ref{eq:FF_stationary1}) as linear combinations of the subtracted form factors $f^{[p]}_\beta(\tau, \partial)$ with some coefficients $c(\tdt)$. This, in addition to numerical structures like (\ref{eq:ibp_r_beta}), leads to new integrals involving the fundamental thermal form factor $ f_\beta(\tau, \partial)$
		\begin{eqnarray}
			&&\int_0^\infty \frac{d\tau}{\tau} \frac1{\tau^{D/2}} \tau^{p} \, c(\tdt) f_\beta(\tau, \partial)\nonumber\\
        &&\qquad\qquad\qquad=c\bigl(\tfrac{D}2-p\bigr)\phi^{(D/2-p)}_{\beta}(\partial).
		\end{eqnarray}
	After the same integration by parts they are expressed in terms of the following {\em basic set of nonlocal operators} $\phi^{(q)}_{\beta}(\partial)$ which, together with $\rho^{(q)}$, will serve as main building blocks of the thermal form factors in the curvature squared part of the effective action,
		\begin{equation} \label{eq:basic_int_nl}
			\phi^{(q)}_{\beta}(\partial) = \int_0^\infty \frac{d\tau}{\tau^{q+1}} \, f_\beta(\tau, \partial).
		\end{equation}

Collecting the vacuum and thermal parts one obtains the total one-loop effective action
	\begin{widetext}
		\begin{equation}\label{eq:ea_ans}
		\begin{aligned}
			-W & =  \frac1{2(4\pi)^{D/2}}\int d^D x \sqrt{g} \biggl\{ \frac{\rho^{(D/2)}}{(g_{\xi\xi})^{D/2}}+
			 \frac{\rho^{(D/2-1)}}{(g_{\xi\xi})^{D/2-1}}\biggl(\frac16 R - P \\&\hspace{3em}+ \frac{1}{6} (D-2)\biggl[\mfrac{R_{\xi\xi}}{g_{\xi\xi}} + \bigl(\nabla_{(\mu}\xi_{\nu)}\bigr)^2 -  \bigl(\nabla_\mu \xi^\mu\bigr)^2 - \xi^\mu \, \nabla_\xi \xi^\nu \nabla_{(\mu}\xi_{\nu)} + \frac32 \xi^\mu \, \nabla^\nu \xi^2 \, \nabla_{(\mu}\xi_{\nu)} \biggr] \\
	        &\hspace{3em}+ \frac{1}{12} D(D-2) \Bigl[ (\nabla_\mu\omega)^2 -2 \xi^\mu \, \nabla^\nu \xi^2 \, \nabla_{(\mu}\xi_{\nu)} + 2\xi^\mu \xi^\nu \nabla_{(\mu}\xi_{\nu)} \, \nabla_\rho\xi^\rho \Bigr]  \biggr)   \\
	        &\hspace{3em}+\frac12\biggl(R^{\mu\nu} \, \phi_1(\nabla)R_{\mu\nu} + R \, \phi_2(\nabla)R + P \, \phi_3(\nabla)R + P \, \phi_4(\nabla)P \\
	        & \hspace{5em}+ R \,\phi_5 (\nabla) \mfrac{\Box}{\Delta} \bR  + P \, \phi_6(\nabla)\mfrac{\Box}{\Delta}\bR + \bR^{\mu\nu} \mfrac{\Box}{\Delta}\phi_7(\nabla)\mfrac{\Box}{\Delta}\bR_{\mu\nu}
	        + \bR \mfrac{\Box}{\Delta}\phi_8(\nabla)\mfrac{\Box}{\Delta}\bR \biggr)\biggr\}+\mathcal{O}[\Re^3]
		\end{aligned}
	\end{equation}
	with the full set of covariantly completed form factors of its curvature squared terms
		\begin{equation} \label{eq:ea_ff_full}
			\begin{aligned}
	        &\phi_I(\nabla) = \phi^{\mathrm{vac}}_I(-\Box)+\phi^\beta_I(\nabla), \quad I=1,\ldots 4,\\
	        &\phi_I(\nabla) = \phi^\beta_I(\nabla), \quad I=5,\ldots 8,
			\end{aligned}
		\end{equation}
where the vacuum form factors $\phi^{\mathrm{vac}}_I(-\Box)$ are given by Eqs.~(\ref{eq:vac_ff_ea}) and the thermal form factors $\phi^\beta_I(\nabla)$ read
			\begin{eqnarray}
				&&\phi^\beta_1(\nabla) = -\mfrac{2D}{\Box^2} \phi_\beta^{(D/2)}(\nabla) + \mfrac1{\Box}  \phi_\beta^{(D/2-1)}(\nabla)
	            + \rho^{(D/2)}\mfrac{2D}{\Box^2}  +\rho^{(D/2-1)}\mfrac{D-3}{3\,\Box}, \label{eq:ea_ff_expl}\\
				&&\phi^\beta_2(\nabla) =\mfrac{D(D+2)}{4\,\Box^2} \phi_\beta^{(D/2)}(\nabla) - \mfrac1{2\,\Box} \phi_\beta^{(D/2-1)}(\nabla)
	            -\rho^{(D/2)}\mfrac{D(D+2)}{4\,\Box^2} -\rho^{(D/2-1)}\mfrac{D^2+2D-12}{24\,\Box},\\
				&&\phi^\beta_3(\nabla) = -\mfrac{D-2}{\Box}\phi_\beta^{(D/2-1)}(\nabla)+\rho^{(D/2-1)}\mfrac{D-2}{\Box},\\
				&&\phi^\beta_4(\nabla) = \phi_\beta^{(D/2-2)}(\nabla),\\
				&&\phi^\beta_5(\nabla) = -\mfrac{D(D+1)}{2\,\Box^2}\phi_\beta^{(D/2)}(\nabla) + \mfrac{D}{4\,\Box}\phi_\beta^{(D/2-1)}(\nabla)
	            + \rho^{(D/2)}\mfrac{D(D+1)}{2\,\Box^2} + \rho^{(D/2-1)}\mfrac{D(D-2)}{12\,\Box}, \\
				&&\phi^\beta_6(\nabla) = \mfrac{D-1}{\Box} \phi_\beta^{(D/2-1)}(\nabla)
	            -  \mfrac{1}{2} \phi_\beta^{(D/2-2)}(\nabla)-\rho^{(D/2-1)}\mfrac{D-1}{\Box},\\
				&&\phi^\beta_7(\nabla) = \mfrac{2(D+1)}{\Box^2}\phi_\beta^{(D/2)}(\nabla)
	            - \mfrac1{\Box}\phi_\beta^{(D/2-1)}(\nabla)-\rho^{(D/2)}\mfrac{2(D+1)}{\Box^2} - \rho^{(D/2-1)}\mfrac{D-2}{3\,\Box},\\
				&&\phi^\beta_8(\nabla) =\mfrac{D^2-1}{4\,\Box^2} \phi_\beta^{(D/2)}(\nabla)\!
	            -\!\mfrac{D-1}{4\Box}\phi_\beta^{(D/2-1)}(\nabla)\!
	            +\!\mfrac1{16}\phi_\beta^{(D/2-2)}(\nabla)
	            \!-\! \rho^{(D/2)}\mfrac{D^2-1}{4\,\Box^2}\! -\! \rho^{(D/2-1)}\mfrac{(D-1)(D-5)}{24\, \Box} \label{eq:ea_ff_exp8}.
			\end{eqnarray}

The calculation of the basic form factor functions $\phi_\beta^{(q)}(\partial)$ arising in the explicit expressions for $\phi_I(\nabla)$, is rather involved and is done in \cite{tbp}. Here we reproduce the answer in the momentum space representation (cf.~$ik_\mu \leftrightarrow \partial_\mu$ and $k_\mu = (k_0, \bk)$) in the following form
		\begin{align}
			&\phi_\beta^{(q)}(ik) = \Bigl(\frac{2}{\beta}\Bigr)^{\!2q} \, k^{2q} \,
		    \partial_{k^2}^{q} \bigl(\varOmega\varPhi_{q}(|k|)\bigr), \label{eq:vphi_ans_k}\\
		    &\varPhi_q(|k|) =  (-)^q\frac{(2\pi)^{2q}}{(2q)!}\frac{4\pi}{i \beta |k|}
	        \Bigl[\zeta'(-2q,-\tfrac{i \beta}{4\pi}|k|)-\zeta'(-2q,\tfrac{i\beta}{4\pi}|k|)\Bigr]
		    - \frac{\pi^2}{(2q)!} \Bigl(\frac{\beta |k|}2\Bigr)^{\!2q-1}\nonumber\\
		    &\qquad\qquad+\sum_{l=0}^q\frac{4}{(2l+1)!}\zeta(2q-2l)\Bigl[\ln\Bigl(\mfrac{\beta |k|}{4\pi}\Bigr)
		    +H_{2q}-H_{2l+1}\Bigr] \Bigl(\frac{\beta |k|}2\Bigr)^{\!2l},  \label{eq:I_p_k}
		    \end{align}
	\end{widetext}
where we denote $|k| = \sqrt{k^2}$.

	The expression involves the Hurwitz zeta function defined as $\zeta(s, a) = \sum_{n=0}^\infty 1/(n+a)^{s}$, and prime denotes the derivative with respect to the first of its arguments, $\zeta'(s, a) = \tfrac{d}{ds}\zeta(s, a)$. Here $\varOmega$ is the translation type operator
	\begin{align}\label{eq:omega_k}
	    	\varOmega\varPhi(|k|) &= \frac{e^{i\vartheta} \varPhi(e^{i\vartheta} |k|) + e^{-i\vartheta}
            \varPhi(e^{-i\vartheta} |k|)}{e^{i\vartheta} + e^{-i\vartheta}}, \\
	    	\vartheta&=\vartheta(|k|,k_0) \equiv \arcsin \frac{k_0}{|k|}. \nonumber
	\end{align}
	defined by its action on a generic test function $\varPhi(|k|)$. Here $\vartheta$ has the meaning of Euclidean rapidity parameter. When calculating the $k^2$-derivatives in (\ref{eq:vphi_ans_k}), the parameter $\vartheta(|k|,k_0) = \arcsin(k_0/|k|)$ should be treated as a function of $|k|=\sqrt{k^2}$ at fixed $k_0$. Note also the structure $e^{\pm i\vartheta} |k| = |\bk| \pm i k_0$, $|\bk| = \sqrt{\bk^2}$, of the test function $\varPhi$ argument.

	Substituting (\ref{eq:omega_k}) and (\ref{eq:I_p_k}) into (\ref{eq:vphi_ans_k}) makes the resulting expression for $\phi^{(p)}_\beta(ik)$ extremely complicated. While the explicit result (\ref{eq:vphi_ans_k}) for the effective action fundamental form factors may be suitable for numerical calculations, it is of a limited use for a further analysis. Therefore, in the next sections we present their high and low temperature expansion and use it in the corresponding limit of the effective action.

\subsection{High temperature expansion}
	Using the series expansion of Hurwitz zeta function, we develop the high-temperature expansion of the form factors (\ref{eq:vphi_ans_k}), which is valid in $\beta\to 0$ limit. The answer reads
	\begin{widetext}
	\begin{align} \label{eq:vphi_beta_ans_k}
		&\phi_\beta^{(q)}(ik)
		= 2{(-k^2)^q}\Biggl[\sum_{l=0}^{q-1} (-)^{l-q}\frac{l!(q-l-1)!}{(2l+1)!}\zeta(2q-2l)\; F_{q,l}(\vartheta) \,
	    \Bigl(\frac{\beta|k|}{2}\Bigr)^{\!2l-2q}
        + \frac{\pi^2}{2^{2q} q!} \Bigl(\frac{k^2}{\bk^2}\Bigr)^q \frac1{\beta|\bk|} \nonumber\\
	    &\hspace{10em}
		+\frac{q!}{(2q+1)!} \Bigl(\log \mfrac{\beta |k|}{4\pi} + \mfrac12 H_q - H_{2q+1} + \gamma \Bigr)
        -\frac{(q+1)!}{(2q+1)!} \, \tilde{F}_q (\vartheta) \\
	    &\hspace{10em}+q!\sum_{l=q+1}^\infty  \frac{(2l-2q)!}{(2l+1)!} \zeta(2l-2q+1) \, C_{2l-2q}^{(q+1)}(\sin \vartheta) \,
	    \Bigl(\frac{\beta|k|}{4\pi}\Bigr)^{\!2l-2q}\Biggr],\nonumber
	\end{align}
	\end{widetext}
	where {$|\bk|=\sqrt{\bk^2}$}, $\gamma$ is the Euler's constant and $C_n^{(\alpha)}(x)$ denotes the Gegenbauer polynomial. Here we also introduce auxiliary functions
		\begin{equation} \label{eq:high_T_aux}
			\begin{aligned}
				&F_{q,l}(\vartheta)= {}_2 F_1(q-l,l+1;\tfrac12;\sin^2\vartheta), \\
				&\tilde{F}_q(\vartheta) = \sin^2\! \vartheta \; {}_3 F_2(1,1,q+2;\tfrac32,2; \sin^2\vartheta),
			\end{aligned}
		\end{equation}
	where the hypergeometric functions ${}_2F_1$ and ${}_3F_2$ reduce to elementary functions for the specific values of their parameters involved. The first line of (\ref{eq:vphi_beta_ans_k}) contains negative powers of $\beta$, i.e.\ terms exhibiting power-law growth in the high-temperature limit. The second line consist of terms involving logarithms and constants in $\beta$. Finally, the third line represents subleading terms, vanishing at $\beta\to 0$. The latter are local, since the combination $C_{2l-2q}^{(q+1)}(k_0/|k|) \, k^{2l}$ contains only {even} positive powers of $k_0$ and $|k|$, {so that apart from the first few nonlocal terms this is a local expansion in spacetime derivatives.} Note that in the static limit, $k_0 \to 0$, non-locality survives only in $\beta^{-1}$-term which confirms the results of \cite{Gusev:1998rp}.
		
		Substituting the expansion (\ref{eq:vphi_beta_ans_k}) into (\ref{eq:ea_ff_expl}) and combining with the vacuum part (\ref{eq:vac_ff_ea}), we obtain the corresponding high-temperature expansion of the effective action (\ref{eq:ea_ans}). As an illustrative and the most physically interesting example, we provide the explicit answer for the form-factors (\ref{eq:ea_ff_full}) in the dimension $D=4$. In this dimension, the auxiliary functions (\ref{eq:high_T_aux}) occur for the following specific set of their parameters,
	\begin{widetext}
		\begin{equation}
		\begin{aligned}
			&F_{1,0}(\vartheta) = \frac{\vartheta  \sin \vartheta +\cos \vartheta }{\cos^2 \vartheta},\quad
F_{2,1}(\vartheta) = F_{2,0}(\sin\vartheta) = \frac{12 \vartheta  \sin \vartheta +9 \cos \vartheta -\cos 3 \vartheta}{8 \cos^5 \vartheta},\\
			&\tilde{F}_{0} (\vartheta) = \frac{\vartheta \sin \vartheta}{\cos \vartheta}, \\
			&\tilde{F}_{1} (\vartheta) = \frac{ 2(3\sin \vartheta + \sin 3\vartheta) \, \vartheta+\cos\vartheta - \cos3\vartheta}{16 \cos^3 \vartheta}, \\
			&\begin{aligned}
				&\tilde{F}_{2} (\vartheta) = \frac1{192 \cos^5 \vartheta}\Bigl(4(10\sin\vartheta
            + 5 \sin 3\vartheta + \sin 5\vartheta) \, \vartheta + 12\cos\vartheta - 9 \cos3\vartheta - 3\cos 5\vartheta\Bigr).
			\end{aligned}
		\end{aligned}\hspace{-2.25em}
		\end{equation}
In terms of them the answer for the form factors (\ref{eq:ea_ff_full}) reads
		\begin{align}
			&\begin{aligned}
				&\phi_1(ik) = -\frac{128 \pi ^4}{45}\frac{F_{2,0}(\vartheta)-1}{\beta^4 k^4} - \frac{4 \pi^2}{9}\frac{3 F_{1,0}(\vartheta)
	        -4 F_{2,1}(\vartheta)+1}{\beta ^2 k^2} - \frac{\pi^2}{2}\frac{\sin^2 \vartheta}{\cos^5\vartheta} \frac1{\beta |k| }
	        +\frac{1}{15}\log\Bigl(\frac{\beta  \mu }{4 \pi}\Bigr)\\
	        &\hspace{4em} + \frac{1}{15} \bigl(\gamma- 10 \tilde{F}_{1}(\vartheta)+ 12 \tilde{F}_{2}(\vartheta)-\tfrac{17}{120}\bigr)+ \frac{\zeta (3)}{1680 \pi ^2} (2 - 5 \cos 2 \vartheta) \, \beta ^2 k^2+\mathcal{O}(\beta ^4),
			\end{aligned}
	        \label{eq:phi_1_ht}\\
			&\begin{aligned}
				&\phi_2(ik) = \frac{32 \pi^4}{15} \frac{F_{2,0}(\vartheta)-1}{\beta^4 k^4} + \frac{2 \pi^2}{3}\frac{F_{1,0}(\vartheta)
	        -2 F_{2,1}(\vartheta)+1}{\beta^2 k^2}-\frac{\pi^2 }{8} \frac{\cos 2 \vartheta -2}{\cos^5 \vartheta }\frac1{\beta |k|}
	        +\frac{1}{30} \log \Bigl(\frac{ \beta  \mu }{4 \pi }\Bigr)\\
	        &\hspace{4em} + \frac{1}{30} \bigl( \gamma+10 \tilde{F}_{1}(\vartheta) - 18 \tilde{F}_{2}(\vartheta)-\tfrac{863}{480} \bigr)+ \frac{\zeta (3)}{3360 \pi ^2} (4-3 \cos 2 \vartheta ) \, \beta^2 k^2+\mathcal{O}(\beta ^4),
			\end{aligned}\\
			&\begin{aligned}
				&\phi_3(ik) = \frac{8 \pi^2}{3}\frac{F_{1,0}(\vartheta)-1}{ \beta ^2 k^2}-\frac{\pi^2 }{\cos^3\vartheta}  \frac1{\beta |k| }
	        -\frac23\log\Bigl(\frac{ \beta  \mu }{4 \pi }\Bigr) \\
	        &\hspace{4em} -\frac{1}{36} (24\gamma-48 \tilde{F}_{1}(\vartheta)-19)+ \frac{\zeta (3)}{120 \pi ^2}(3 \cos 2 \vartheta - 2) \, \beta ^2 k^2+\mathcal{O}(\beta ^4),
			\end{aligned} \\
			&\begin{aligned}
	        &\phi_4(ik) = \frac{2 \pi ^2 }{\cos \vartheta}  \frac1{\beta |k|} +2 \log\Bigl(\frac{ \beta  \mu }{4 \pi }\Bigr)
	        + \bigl(2 \gamma-2 \tilde{F}_{0}(\vartheta)-1\bigr)-\frac{\zeta (3) }{24 \pi ^2} (2 \cos 2 \vartheta -1)\, \beta^2 k^2 +\mathcal{O}(\beta ^4),
	        \end{aligned}\\
			&\begin{aligned}
				&\phi_5(ik) = - \frac{32 \pi^4}{9} \frac{F_{2,0}(\vartheta)-1}{\beta^4 k^4}-\frac{4 \pi^2}{9} \frac{3 F_{1,0}(\vartheta)
	        -5 F_{2,1}(\vartheta)+2}{\beta ^2 k^2}+\frac{\pi^2}{8}\frac{2 \cos 2 \vartheta - 3}{\cos^5\vartheta}  \frac1{\beta |k|}\\
	        &\hspace{4em} -\frac{2 }{3}\tilde{F}_{1}(\vartheta)+\tilde{F}_{2}(\vartheta)+\frac{1}{15}
	        - \frac{\zeta(3)}{1680 \pi ^2} (\cos 2 \vartheta + 1) \, \beta ^2 k^2+\mathcal{O}(\beta ^4),
			\end{aligned}
	        \label{eq:phi_5_ht}\\
			&\begin{aligned}
				&\phi_6(ik) = -4 \pi ^2\frac{F_{1,0}(\vartheta)-1}{\beta^2 k^2}-\frac{\pi^2}{2} \frac{\cos 2 \vartheta - 2}{\cos^3\vartheta} \frac1{\beta |k| }\\
	        &\hspace{4em}+\tilde{F}_{0}(\vartheta)-2 \tilde{F}_{1}(\vartheta)-\frac{1}{3}+\frac{\zeta (3)}{240 \pi ^2} (\cos 2 \vartheta +1)\, \beta^2 k^2+\mathcal{O}(\beta ^4),
			\end{aligned}\\
			&\begin{aligned}
				&\phi_7(ik) = \frac{32 \pi^4}{9} \frac{F_{2,0}(\vartheta)-1}{\beta^4 k^4} + \frac{4 \pi^2}{9} \frac{3 F_{1,0}(\vartheta)
	        -5 F_{2,1}(\vartheta)+2}{\beta^2 k^2} + \frac{\pi^2}{8} \frac{4 \sin ^2(\vartheta )+1}{\cos^5 \vartheta}  \frac1{\beta |k| }\\
	        &\hspace{4em}+\frac{2 }{3}\tilde{F}_{1}(\vartheta)-\tilde{F}_{2}(\vartheta)-\frac{1}{15}+\frac{\zeta (3) }{1680 \pi ^2} (\cos2\vartheta + 1) \beta^2 k^2 +\mathcal{O}(\beta ^4),
			\end{aligned}\\
			&\begin{aligned}
				&\phi_8(ik) = \frac{4 \pi^4}{3} \frac{F_{2,0}(\vartheta)-1}{\beta^4 k^4}+ \frac{\pi^2}{6} \frac{6 F_{1,0}(\vartheta)
	            -5 F_{2,1}(\vartheta)-1}{\beta ^2 k^2}+ \frac{\pi^2}{64} \frac{ -8 \cos 2 \vartheta+\cos4 \vartheta+6}{\cos^5\vartheta} \frac{1}{\beta |k|}\\
	        &\hspace{4em}+\frac{1}{120} (-15 \tilde{F}_{0}(\vartheta)+60 \tilde{F}_{1}(\vartheta)-45 \tilde{F}_{2}(\vartheta)+2)-\frac{\zeta (3)}{3360 \pi ^2} (\cos2\vartheta + 1) \, \beta^2 k^2 +\mathcal{O}(\beta ^4).
			\end{aligned}
	\label{eq:phi_8_ht}
		\end{align}
	\end{widetext}
	In these expressions each term consists of a factor depending on $\vartheta(|k|,k_0) = \arcsin(k_0/|k|)$ multiplied by a power of the dimensionless combination $\beta |k|$. The form factors $\phi_1,\dots,\phi_4$ contain logarithmic contributions of the form $\log (\beta\mu)$, whereas $\phi_5,\dots,\phi_8$ are free from logarithms.
			
	This is a general pattern of high-temperature behavior---when a form factor includes both vacuum and purely thermal contributions, the vacuum part contains logarithms $\log (k^2/\mu^2)$ while the thermal part contains logarithms $\log(\beta |k|)$. The $k$-dependence cancels exactly between them, leaving a net logarithmic dependence in the form of
	    \begin{equation}
	    \log(\beta\mu) = -\tfrac12\log(k^2/\mu^2) + \log(\beta |k|).
	    \end{equation}
	{This, in its turn, follows from the fact that the basic vacuum and thermal form factors $\phi^{(q)}_{\rm vac}(-\Box)$ and $\phi^{(q)}_\beta(\nabla)$, defined respectively by Eqs.~(\ref{eq:vphi_vac_ans_k}) and (\ref{eq:basic_int_nl}), additively enter the full formfactor (\ref{eq:ea_ff_full}). Their $\tfrac12\log(-\Box)=\log|k|$ terms differ only by sign (compare (\ref{eq:vphi_vac_ans_k}) with the high-temperature limit (\ref{eq:vphi_beta_ans_k}) of the thermal $\phi^{(q)}_\beta$) and, therefore, completely cancel out.} Consequently, the form factors $\phi_1,\dots,\phi_4$, which have a nonvanishing vacuum part, exhibit $\log(\beta\mu) $ terms.
		
	For $\phi_5,\dots,\phi_8$ the situation is different because they have no vacuum part. If their thermal contributions could have produced logarithms, those logarithms would be of the form $\log(\beta k)$ and could not be canceled by vacuum logarithms. The fact that $\phi_5,\dots,\phi_8$ are completely free of logarithms, therefore, implies that any $\log(\beta |k|)$ terms within their purely thermal sector get completely cancelled among themselves. This is clearly seen in the expressions (\ref{eq:phi_5_ht})--(\ref{eq:phi_8_ht}).
		
	It is instructive to compare the results (\ref{eq:phi_1_ht})--(\ref{eq:phi_8_ht}) for the form factors with the stationary case, previously studied in \cite{Gusev:1998rp,Valle:2025fev}. The stationary limit corresponds to setting $k_0 = 0$ (equivalently $\vartheta = 0$). As can be seen from (\ref{eq:phi_1_ht})--(\ref{eq:phi_8_ht}) together with $F_{q,l}(0) = \tilde{F}_q(0) = 1$, the leading $\beta^{-4}$ and $\beta^{-2}$ terms in these form factors then vanish. Consequently, in the stationary limit the dominant high-temperature contributions come from the terms proportional to the zeroth and first powers of the curvature (the first three lines of (\ref{eq:ea_ans})). These structures are local and scale as $\beta^{-4}$ and $\beta^{-2}$, respectively, in $D=4$. The subdominant contribution arises from the $\beta^{-1}$ terms of (\ref{eq:phi_1_ht})--(\ref{eq:phi_8_ht}) that survive the stationary limit. Such terms were first observed in \cite{Gusev:1998rp} and correspond to a structure of the form $\Re \tfrac{1}{\beta\sqrt{-\Delta}}\Re$ in the effective action.
		
	These observations extend to general even dimensions $D$ with only minor modifications. First, the cancellation of $\log|k|$ terms takes place in any even dimension. In the stationary limit, the dominant terms still come from the zeroth and first orders in the curvature, scaling respectively as $\beta^{-D}$ and $\beta^{-D+2}$. The curvature-squared part of the effective action admits a high-temperature expansion which begins with terms of order $\beta^{-D}$ and $\beta^{-D+2}$. The $\beta^{-D}$ and $\beta^{-D+2}$ terms vanish in the stationary limit, so for even $D \ge 6$ the subdominant contribution from the curvature-squared sector is of the form $\beta^{-D+4}\,\Re^2$ (which is local in contrast to $D=4$ case). All these statements directly follow by substituting (\ref{eq:vphi_beta_ans_k}) into (\ref{eq:ea_ff_expl}) and adding the vacuum part (\ref{eq:vac_ff_ea}).

\subsection{Low-temperature expansion}
	Along with the high-temperature limit, one can also build a low-temperature, large $\beta$, expansion of the effective action form factors. The fundamental purely thermal form factor has the asymptotic series
	\begin{align}
			\phi_\beta^{(q)}(k)\sim 4 q! \, \Bigl(\frac{2}{\beta}\Bigr)^{\!2q}&\sum_{l=0}^\infty  (-)^l (2l)! \, \zeta(2(l+q+1))
            \nonumber\\ & \times \,  C_{2l}^{(q+1)}\bigl(\mfrac{k_0}{|k|}\bigr) \, \Bigl(\frac2{\beta |k|}\Bigr)^{\!2l+2},
	\end{align}
which, in contrast to its high-temperature counterpart (\ref{eq:vphi_beta_ans_k}), is a pure power-law expansion. The substitution of this series into (\ref{eq:ea_ff_expl}) yields the low-temperature expansion of the form factors. Purely thermal contributions to these form factors exhibit a universal leading-order scaling $\beta^{-D+2} = T^{D-2}$. Consequently, in the non-local expansion for even dimensions $D \ge 4$, purely thermal form factors are subdominant, and the leading-order contribution to the total form factor (\ref{eq:ea_ff_full}) comes from its vacuum part (\ref{eq:vac_ff_ea}).

\section{Discussion} \label{sec:discussion}

	Main results of this work comprise the curvature expansion for the heat kernel trace (\ref{eq:TrK_cov}) and the one-loop effective action (\ref{eq:ea_ans}) with the explicit expressions for the gravitational form factors {of the latter (\ref{eq:ea_ff_expl})--(\ref{eq:ea_ff_exp8}), built in terms of the basic operator functions (\ref{eq:vphi_ans_k})--(\ref{eq:omega_k}) and numerical coefficients (\ref{eq:rho_def}). Their high ``temperature'' asymptotic behavior in the quasithermal setup of a nonvacuum quantum state is given by Eqs.~(\ref{eq:phi_1_ht})--(\ref{eq:phi_8_ht}).} This setup implies a non-static and non-stationary Euclidean background spacetime with periodic boundary conditions of the period $\beta=1/T$ where $T$ plays the role of effective global temperature to be locally modified by the metric gravitational potential. The results are obtained in the approximation quadratic in metric perturbations on top of flat Euclidean space and covariantized, or in the language of \cite{Donoghue:2015xla} nonlinearly completed, in terms of spacetime curvature. Covariantization includes a special vector field $\xi^\mu(x)$ which generalizes the Killing vector of static geometries with time translation isometry to the case of a generic arbitrarily inhomogeneous metric subject to a single timelike periodicity condition. This vector field is obtained as a covariant metric functional to quadratic order in metric perturbations and gives rise to the local function $T/\sqrt{\xi^2(x)}$, $\xi^2(x)=g_{\mu\nu}(x)\xi^\mu(x)\xi^\nu(x)$, reducing to Tolman temperature $T/\sqrt{g_{00}(x)}$ on stationary manifolds with Killing vector $\xi^\mu(x)=\delta^\mu_0$.
		
	Obtained algorithms look frustratingly complicated---addition of a single dimensional parameter $\beta$ with the requirement of periodic boundary conditions immediately gives rise to numerous new tensor structures and significantly extends the set of nonlocal form factors and their complexity. At the same time, this is only a small piece of the total effective action, quadratic in curvatures, not to mention that this is just one-loop approximation. Moreover, it is restricted to the case of a very simple topology of a spatial slice of spacetime, which is supposed to be asymptotically flat in spatial directions. Though potentially the above results in view of their universality can be meaningful in various applications, it is worth pointing out a concrete physical setup in which all this formalism can be directly useful despite all its perturbation theory limitations.
		
	Another objection, that can be raised against the usefulness of this universal algorithm and its covariantization, is the complexity of nonlocal form factor functions of the {\em exact} curved-space d'Alembertian or Laplacian. Their feasibility can be called in question as long as their implementation is not available in exact form unlike their flat-space counterparts. For the perturbation scheme to be feasible and practicable, all ingredients of its leading order should be exactly available in analytical form---the problem raised in \cite{Donoghue:2015xla}. This problem is not at all compelling for local differential operator structures, but may be very stringent for nonlocal form factors whose exact kernels are not known. Various approaches to this difficulty are conceivable, ranging from the search for special representations of these form factors, like spectral representations of \cite{Barvinsky:1990uq,Vilkovisky:2007ny}, to a possible retreat back to their analytical flat space form.
	
	With regard to the above mentioned limitation of the small curvature and one-loop approximation we can point out a concrete model that served as a main motivation for this work. This is a model of initial conditions for inflationary cosmology in the form of the microcanonical density matrix \cite{Barvinsky:2007vb} represented by the Euclidean path integral over periodic configurations. When the model is dominated by numerous conformally invariant fields it incorporates the quasi-thermal stage preceding inflation \cite{Barvinsky:2006uh} and provides the origin of the Higgs-type or Starobinsky $R^2$-type inflationary scenario \cite{Starobinsky:1980te,Barvinsky:2015uxa}. This scenario turns out to be a subplanckian phenomenon subject to semiclassical $1/N$-perturbation theory in the number of higher spin conformal fields \cite{Barvinsky:2015wvz}, which justifies the use of curvature expansion.
	
	Thus, the above results on the effective action for nonvacuum states become critically important for the studies of this model, especially for its CMBR spectrum, in particular, because the quadratic order in curvatures is responsible for the propagator of cosmological perturbations. It should be emphasized that the flat space background of our nonvacuum perturbation theory does not contradict the cosmological Friedmann metric background, because the conformal mode of the Friedmann metric is treated in \cite{Barvinsky:2006uh,Barvinsky:2015uxa} by the nonperturbative method of conformal transformation from the Friedmann universe to asymptotically flat spacetime. The contribution of this transformation for conformal fields is exactly calculable via their one-loop trace anomaly, and for higher spin conformal fields, due to their higher-derivative superrenormalizable nature, Weyl anomaly is restricted to the one-loop order of the semiclassical expansion. Similar mechanisms justifying the validity of one-loop results work also within the scope of recently suggested models of infinite-derivative quantum gravity with nonlocal form factors based on entire functions \cite{Buoninfante:2018xiw,Koshelev:2017ebj,Koshelev:2025pxg}---UV divergences in these models can also be restricted to the one-loop order approximation.
	
	Direct physical predictions following from the structure of the obtained effective action go beyond the material of this paper. However, already at this stage one can infer some information on the form of its quasi-thermal corrections to gravitational equations. In particular, zero order curvature corrections represent a kind of a radiation component, because the metric variation of the first term of (\ref{eq:TrK_cov}) is traceless. Terms linear in the curvature are expected to be anisotropic in spacetime because of the presence of the vector $\xi^\mu$, which means that the thermally corrected effective gravitational coupling ``constant'' becomes anisotropic and, moreover, nonlocal tensor, etc. Details of these modifications will be considered in future publications.
	
	To accomplish this paper let us briefly dwell on the list of problems subject to further studies.  One of them is unclear geometrical meaning of the vector field $\xi^\mu$ generalizing the notion of the Killing isometry to generic inhomogeneous metrics. This includes the task of clarifying the two-parameter ambiguity in its construction, the behavior of its integral curves and the relation to the geometrical meaning of the period parameter $\beta$.
	
	The second most important task is the construction of the Schwinger-Keldysh generating functional from the obtained Euclidean effective action. This corresponds to the deformation of the Euclidean closed contour into the complex plane of time, where the Lorentzian segments of this contour are associated with the unitary back and forth evolution in physical time---bras and kets of the path integral \cite{Witten:2025ayw,Fumagalli:2024msi}. From \cite{Barvinsky:2023jkl} it is known that various Green's functions of Schwinger-Keldysh diagrammatic technique for the density matrix of the system prescribed by the Euclidean path integral can be obtained by analytical continuation from the purely Euclidean contour and, moreover, satisfy the famous KMS condition (contrary to the observer's no-boundary states of \cite{Blommaert:2025bgd}). However, the resulting physical field correlators on these Lorentzian segments hardly obey the same analytical continuation rule for a generic inhomogeneous geometry, in contrast with the homogeneous de Sitter space where this continuation seems possible \cite{Higuchi:2010xt}. Probably, the needed analytical rule is more complicated and includes some extra causality retardation prescription like the one observed for vacuum setup in asymptotically flat spacetime \cite{Barvinsky:1987uw}.

\section*{Acknowledgments}
	The authors thank “BASIS” Foundation for the Advancement of Theoretical Physics and Mathematics for support. The authors are also grateful to D. Diakonov, I.G. Pirozhenko, and D. Sadekov for useful discussions.

\appendix

\section{Basis tensors} \label{sec:app_basis_tens}

	This Appendix lists the basis tensors used in the expansion of the second‑order heat‑kernel coefficient $\Tr K_2(\tau)$. These tensors appear in the ansatz (\ref{eq:Pi_ansatz}) for the operators $\varPi_{\mu\nu,\rho\sigma}$ and $\varPi_{\mu\nu}$ defined in Eq.~(\ref{eq:TrK2}).

	The basis of fourth rank tensors constructed out of $\eta_{\mu\nu}$, $k_\mu$ and $\bar{\xi}^\mu$ and having the symmetries $\mu \leftrightarrow \nu$, $\rho \leftrightarrow \sigma$ and $(\mu,\nu) \leftrightarrow (\rho,\sigma)$ contains the following 14 elements
	\begin{equation*} 
		\begin{aligned}
			&(\varPi_1)_{\mu\nu,\rho\sigma} = \eta_{\mu\rho}\eta_{\nu\sigma} + \eta_{\mu\sigma}\eta_{\nu\rho}, \\	
			&(\varPi_2)_{\mu\nu,\rho\sigma} = \eta_{\mu\rho}\bar{\xi}_\nu \bar{\xi}_\sigma + \eta_{\mu\sigma}\bar{\xi}_\nu \bar{\xi}_\rho + \eta_{\nu\rho}\bar{\xi}_\mu \bar{\xi}_\sigma + \eta_{\nu\sigma}\bar{\xi}_\mu \bar{\xi}_\rho, \\
			&(\varPi_3)_{\mu\nu,\rho\sigma} = \bar{\xi}_\mu \bar{\xi}_\nu \bar{\xi}_\rho \bar{\xi}_\sigma,\\
			&(\varPi_4)_{\mu\nu,\rho\sigma} = \eta_{\mu\nu} \eta_{\rho\sigma},\\
			&(\varPi_5)_{\mu\nu,\rho\sigma} = \bar{\xi}_\mu \bar{\xi}_\nu \eta_{\rho\sigma} + \bar{\xi}_\rho \bar{\xi}_\sigma \eta_{\mu\nu},\\
		\end{aligned}
	\end{equation*}
	\begin{equation*} 
		\begin{aligned}
			&\begin{aligned}
				(\varPi_6)_{\mu\nu,\rho\sigma} = \bar{\xi}_\mu k_\rho \eta_{\nu\sigma} + \bar{\xi}_\rho k_\mu \eta_{\nu\sigma} + \bar{\xi}_\mu k_\sigma \eta_{\nu\rho} + \bar{\xi}_\sigma k_\mu \eta_{\nu\rho}& \\+\bar{\xi}_\nu k_\rho \eta_{\mu\sigma} + \bar{\xi}_\rho k_\nu \eta_{\mu\sigma} + \bar{\xi}_\nu k_\sigma \eta_{\mu\rho} + k_\sigma k_\nu \eta_{\mu\rho}&,
			\end{aligned}\\
			&(\varPi_7)_{\mu\nu,\rho\sigma} = k_\mu \bar{\xi}_\nu \bar{\xi}_\rho \bar{\xi}_\sigma {+} k_\nu \bar{\xi}_\mu \bar{\xi}_\rho \bar{\xi}_\sigma {+} k_\rho \bar{\xi}_\mu \bar{\xi}_\nu \bar{\xi}_\sigma {+} k_\sigma \bar{\xi}_\mu \bar{\xi}_\nu \bar{\xi}_\rho, \\
			&(\varPi_8)_{\mu\nu,\rho\sigma} = k_\mu k_\rho \eta_{\nu\sigma} + k_\mu k_\sigma \eta_{\nu\rho} + k_\nu k_\rho \eta_{\mu\sigma} + k_\nu k_\sigma \eta_{\mu\rho}, \\
			&(\varPi_9)_{\mu\nu,\rho\sigma} = k_\mu k_\rho \bar{\xi}_\nu \bar{\xi}_\sigma {+} k_\mu k_\sigma \bar{\xi}_\nu \bar{\xi}_\rho {+} k_\nu k_\rho \bar{\xi}_\mu\bar{\xi}_\sigma {+} k_\nu k_\sigma \bar{\xi}_\mu\bar{\xi}_\rho, \\
			&(\varPi_{10})_{\mu\nu,\rho\sigma} = k_\mu k_\nu \bar{\xi}_\rho \bar{\xi}_\sigma + k_\rho k_\sigma \bar{\xi}_\mu \bar{\xi}_\nu,\\
			&(\varPi_{11})_{\mu\nu,\rho\sigma} = \bar{\xi}_\mu k_\nu k_\rho k_\sigma {+} \bar{\xi}_\nu k_\mu k_\rho k_\sigma {+} \bar{\xi}_\rho k_\mu k_\nu k_\sigma {+} \bar{\xi}_\sigma k_\mu k_\nu k_\rho,\\
			&(\varPi_{12})_{\mu\nu,\rho\sigma} = k_\mu k_\nu k_\rho k_\sigma, \\
			&(\varPi_{13})_{\mu\nu,\rho\sigma} = k_\mu k_\nu \eta_{\rho\sigma} + k_\rho k_\sigma \eta_{\mu\nu}, \\
			&(\varPi_{14})_{\mu\nu,\rho\sigma} = k_\mu \bar{\xi}_\nu \eta_{\rho\sigma} + k_\nu \bar{\xi}_\mu \eta_{\rho\sigma} + k_\rho \bar{\xi}_\sigma \eta_{\mu\nu} + k_\sigma \bar{\xi}_\rho \eta_{\mu\nu}.
		\end{aligned}
	\end{equation*}
	
In turn, the basis of second rank symmetric tensors built in terms of the metric $\eta_{\mu\nu}$ and the vectors $k_\mu$ and $\bar\xi_\mu$ consists of
	\begin{equation*}
		\begin{aligned}
			&(\varPi_{15})_{\mu\nu} = \eta_{\mu\nu}, \\
			&(\varPi_{16})_{\mu\nu} = \bar{\xi}_\mu \bar{\xi}_\nu, \\
			&(\varPi_{17})_{\mu\nu} = k_\mu k_\nu, \\
			&(\varPi_{18})_{\mu\nu} = k_\mu \bar{\xi}_\nu + k_\nu \bar{\xi}_\mu.
		\end{aligned}
	\end{equation*}

\section{Explicit form of coefficient functions} \label{sec:app_cfs_ans}
The following expressions give the momentum-space form of the coefficient functions $C_a$, $a=1,\ldots18$, which enter the ansatz (\ref{eq:Pi_ansatz}) for the quadratic part of the heat kernel. They are derived from the perturbative expansion of Sect.~\ref{sec:heat_kernel_h} as functions of $k_\mu = (k_0, \bk)$.
\begin{widetext}
	\begin{align*}
		&C_1 = \frac14 f - \frac12 r, \\
		&C_2 = -\frac18\frac{k^2}{\bk^2}\biggl[ 2(2\tdt+1) + \tau k^2 \biggr]f + \frac14 \biggl[2\tdt + \frac{k^2}{\bk^2}\biggr]r, \\
		&\begin{aligned}
			C_3 =  \frac1{16}\frac{k^4}{\bk^4}\biggl[4(2\tdt + 1)(2\tdt+3) + 4\tau k^2 (2\tdt+3) + (\tau k^2)^2 \biggr] f - \frac18\biggl[4 \frac{k^2}{\bk^2} \tdt +2\frac{k^4}{\bk^4} (2\tdt+3) +  \tau k^2 \frac{k^4}{\bk^4} \biggr] r,
		\end{aligned} \\
		&C_4 = \frac1{16}\Bigl[4 + 4\tau k^2 + (\tau k^2)^2\Bigr] f - \frac18 \tau k^2 r,\\
&\begin{aligned}
			C_5 = -\frac1{16}\frac{k^2}{\bk^2}\biggl[4(2\tdt+1) + 4\tau k^2 (\tdt + 1) + (\tau k^2)^2\biggr] f+
			\frac18\frac{k^2}{\bk^2}\Bigl[2 + \tau k^2 \Bigr] r,
		\end{aligned}\\
&C_6 = \frac18 \frac{k_0}{\bk^2}\biggl[2(2\tdt+1)+\tau k^2\biggr]f - \frac14  \frac{k_0}{\bk^2} r,\\
		&\begin{aligned}
			C_7 = -\frac1{16} k_0\frac{k^2}{\bk^4}\biggl[ 4(2\tdt+3)(2\tdt+1) +4\tau k^2 (2\tdt+3) + (\tau k^2)^2\biggr] f +\frac18\frac{k_0}{\bk^2}\biggl[4\tdt+2\frac{k^2}{\bk^2}(2\tdt+3) + \tau k^2 \frac{k^2}{\bk^2} \biggr]r,
		\end{aligned} \\
		&C_8 = \frac18 \frac1{k^2}\biggl[4\tdt-2\frac{k^2}{\bk^2}(2\tdt+1) + \tau k^2 \Bigl(1-\frac{k^2}{\bk^2}\Bigr)\biggr]f + \frac14 \frac1{\bk^2} r,\\
		&\begin{aligned}
			C_9 = \frac1{16}\frac1{\bk^2}&\biggl[4\Bigl(-2(\tdt+1)+\frac{k^2}{\bk^2}(2\tdt+3)\Bigr)(2\tdt+1)+2\tau k^2 \Bigl(-(4\tdt+5)+2\frac{k^2}{\bk^2}(2\tdt+3) \Bigr) + (\tau k^2)^2 \Bigl(\frac{k^2}{\bk^2} - 1\Bigr)
			\biggr] f \\+
			\frac18 \frac1{\bk^2}&\biggl[4-2\frac{k^2}{\bk^2}(2\tdt + 3) + \tau k^2 \Bigl(-\frac{k^2}{\bk^2}+1\Bigr)\biggr]r,
		\end{aligned} \\
		&\begin{aligned}
			C_{10} = \frac1{16}\frac1{\bk^2}&\biggl[4\Bigl(-2(\tdt+1)+\frac{k^2}{\bk^2}(2\tdt+3)\Bigr)(2\tdt+1)+ 4\tau k^2\Bigl(-(\tdt+2)+\frac{k^2}{\bk^2}(2\tdt+3)\Bigr)+(\tau k^2)^2\frac{k^2}{\bk^2}\biggr]f \\+
			\frac18 \frac1{\bk^2}&\biggl[4-2\frac{k^2}{\bk^2}(2\tdt + 3) - \tau k^2 \frac{k^2}{\bk^2}\biggr]r,		
		\end{aligned}\\
&\begin{aligned}
			C_{11} = \frac1{16}\frac{k_0}{k^4}&\biggl[-4\Bigl(\frac{k^4}{\bk^4} (2\tdt+3)-2 \frac{k^2}{\bk^2} \tdt\Bigr)(2\tdt+1) + 4\tau k^2 \Bigl(\frac{k^2}{\bk^2} (\tdt+1)-\frac{k^4}{\bk^4} (2\tdt+3)\Bigr)-(\tau k^2)^2\frac{k^4}{\bk^4}\biggr]f \\
			+\frac18\frac{k_0}{\bk^4}&\biggl[2(2 \tdt+3) + \tau k^2\biggr]r,
		\end{aligned}\\
&\begin{aligned}
			C_{12} = \frac1{16} \frac1{k^4} \biggl[& 4 \Bigl(4\tdt(\tdt-1)-4\frac{k^2}{\bk^2}\tdt(2\tdt+1)+\frac{k^4}{\bk^4}(2\tdt+1)(2\tdt+3)\Bigr)
\\ &+ 4\tau k^2\Bigl(-2\frac{k^2}{\bk^2}(\tdt+1) +\frac{k^4}{\bk^4}(2\tdt+3) \Bigr) + (\tau k^2)^2 \frac{k^4}{\bk^4}\biggr]f+\frac18\frac1{k^4}\biggl[ 4\frac{k^2}{\bk^2}\tdt - 2\frac{k^4}{\bk^4}(2\tdt+3) - \tau k^2 \frac{k^4}{\bk^4} \biggr]r,
		\end{aligned}\\
	\end{align*}
	\begin{align*}
		&\begin{aligned}
			C_{13} = \frac1{16} \frac1{k^2} \biggl[8\tdt - 4\frac{k^2}{\bk^2}(2\tdt+1)  + 4\tau k^2 \Bigl(\tdt-\frac{k^2}{\bk^2}(\tdt+1)\Bigr)- (\tau k^2)^2 \frac{k^2}{\bk^2}\biggr]f +
			\frac18 \frac{1}{\bk^2} \Bigl[2 + \tau k^2\Bigr] r ,
		\end{aligned}\\
            &C_{14} = \frac1{16}\frac{k_0}{\bk^2} \biggl[4(2\tdt+1) + 4\tau k^2 (\tdt+1) + (\tau k^2)^2\biggr] f - \frac18 \frac{k_0}{\bk^2}\Bigl[2 + \tau k^2\Bigr] r,\\
            &C_{15} = -\frac14 \frac1{k^2}\Bigl[2 \tau k^2 + (\tau k^2)^2\Bigr] f, \\
		&C_{16} =\frac14 \frac1{\bk^2} \Bigl[2\tau k^2(1+2\tdt) + (\tau k^2)^2\Bigr]f -\frac12\frac1{\bk^2} \tau k^2 \, r, \\
		&C_{17} = -\frac14 \frac1{k^4} \biggl[\tau k^2\Bigl(4\tdt- 2\frac{k^2}{\bk^2}(2\tdt+1)\Bigr) - (\tau k^2)^2\frac{k^2}{\bk^2}\biggr]f -\frac12\frac1{k^2\bk^2} \tau k^2 \, r ,\\
		&C_{18} = -\frac14\frac{k_0}{k^2\bk^2} \Bigl[2\tau k^2 (2\tdt+1) + (\tau k^2)^2\Bigr]f + \frac12 \frac{k_0}{k^2\bk^2} \tau k^2 \,r. \\
	\end{align*}
\end{widetext}

\section{Explicit form of $\varXi^0_{\mu\nu,\rho\sigma}$}
\label{sec:app_Xi4}
The tensor $\varXi^0_{\mu\nu,\rho\sigma}(ik, -ik)$, which determines the quadratic part of the temporal component of the quasi-Killing vector $\xi^\mu$ in (\ref{eq:xi}) and which is built out of the objects $k_\mu$, $\bar{\xi}_\mu$ and $\eta_{\mu\nu}$, has the following decomposition in the basis of Appendix \ref{sec:app_basis_tens}
\begin{equation}
	\varXi^0_{\mu\nu,\rho\sigma}(ik, -ik) = \sum_{a=1}^{14} X_a \, (\varPi_a)_{\mu\nu,\rho\sigma}
\end{equation}
with the coefficients
\begin{align*}
	&X_1 = 0,\\
	&X_2 = \frac{k_0^2 k^2}{4\bk^4},\\
	&X_3 = -\frac{k_0^2 k^2}{4 \bk^6} \bigl(a(a+4)k_0^2 + (a(a+6)+2)\bk^2\bigr),\\
	&X_4 = -\frac{k_0^2}{4\bk^2}b^2, \\
	&X_5 = \frac{k_0^2}{4\bk^4}b \, \bigl((a + 2)k_0^2 + (a+3)\bk^2\bigr),
\end{align*}
\begin{align*}
	&X_6 = \frac{k_0^3}{4\bk^4},\\
	&X_7 = \frac{k_0}{4\bk^6} \bigl(a(a+4)k_0^4 + (a^2+6a+2)k_0^2\bk^2 + (a+1)\bk^4\bigr), \\
	&X_8 = -\frac{k_0^4}{4\bk^4 k^2}, \\
	&X_9 = -\frac{k_0^2}{4\bk^6}\bigl(a(a+4)k_0^2 + (2a + 1)\bk^2\bigr),\\
	&\begin{aligned}
		&X_{10} = -\frac{k_0^2}{4 \bk^6 k^2} \bigl(a(a{+}4)k_0^4 + (a^2{+}(b{+}6)a{+}2(b{+}1))k_0^2 \bk^2 \\ & \hspace{7em} + ((b+1)a+3b+2)\bk^4\bigr),		\end{aligned}\\
	&X_{11} = \frac{k_0}{4\bk^6 k^2}\bigl(a(a{+}4)k_0^4 + ((b{+}2)a{+}2b)k_0^2\bk^2 {+} (b{+}1)\bk^4\bigr),\\
	&\begin{aligned}
		&X_{12} = -\frac{k_0^2}{4\bk^6 k^4} \bigl(a(a+4)k_0^4 + 2((b+1)a+2b-1)k_0^2 \bk^2  \\ &\hspace{7em} + (b^2+2b+2)\bk^4\bigr),
	\end{aligned}\\
	&X_{13} = \frac{k_0^2}{4 \bk^4 k^2} b\, \bigl((a+2)k_0^2 + (b+1) \bk^2\bigr),\\
	&X_{14} = -\frac{k_0}{4\bk^4}b \,\bigl((a+2)k_0^2+ \bk^2\bigr).
\end{align*}

\bibliography{main}

\begin{thebibliography}{43}%
\makeatletter
\providecommand \@ifxundefined [1]{%
 \@ifx{#1\undefined}
}%
\providecommand \@ifnum [1]{%
 \ifnum #1\expandafter \@firstoftwo
 \else \expandafter \@secondoftwo
 \fi
}%
\providecommand \@ifx [1]{%
 \ifx #1\expandafter \@firstoftwo
 \else \expandafter \@secondoftwo
 \fi
}%
\providecommand \natexlab [1]{#1}%
\providecommand \enquote  [1]{``#1''}%
\providecommand \bibnamefont  [1]{#1}%
\providecommand \bibfnamefont [1]{#1}%
\providecommand \citenamefont [1]{#1}%
\providecommand \href@noop [0]{\@secondoftwo}%
\providecommand \href [0]{\begingroup \@sanitize@url \@href}%
\providecommand \@href[1]{\@@startlink{#1}\@@href}%
\providecommand \@@href[1]{\endgroup#1\@@endlink}%
\providecommand \@sanitize@url [0]{\catcode `\\12\catcode `\$12\catcode
  `\&12\catcode `\#12\catcode `\^12\catcode `\_12\catcode `\%12\relax}%
\providecommand \@@startlink[1]{}%
\providecommand \@@endlink[0]{}%
\providecommand \url  [0]{\begingroup\@sanitize@url \@url }%
\providecommand \@url [1]{\endgroup\@href {#1}{\urlprefix }}%
\providecommand \urlprefix  [0]{URL }%
\providecommand \Eprint [0]{\href }%
\providecommand \doibase [0]{https://doi.org/}%
\providecommand \selectlanguage [0]{\@gobble}%
\providecommand \bibinfo  [0]{\@secondoftwo}%
\providecommand \bibfield  [0]{\@secondoftwo}%
\providecommand \translation [1]{[#1]}%
\providecommand \BibitemOpen [0]{}%
\providecommand \bibitemStop [0]{}%
\providecommand \bibitemNoStop [0]{.\EOS\space}%
\providecommand \EOS [0]{\spacefactor3000\relax}%
\providecommand \BibitemShut  [1]{\csname bibitem#1\endcsname}%
\let\auto@bib@innerbib\@empty
\bibitem [{\citenamefont {Barvinsky}\ and\ \citenamefont
  {Vilkovisky}(1987)}]{Barvinsky:1987uw}%
  \BibitemOpen
  \bibfield  {author} {\bibinfo {author} {\bibfnamefont {A.~O.}\ \bibnamefont
  {Barvinsky}}\ and\ \bibinfo {author} {\bibfnamefont {G.~A.}\ \bibnamefont
  {Vilkovisky}},\ }\bibfield  {title} {\bibinfo {title} {{Beyond the
  Schwinger-Dewitt Technique: Converting Loops Into Trees and In-In
  Currents}},\ }\href {https://doi.org/10.1016/0550-3213(87)90681-X} {\bibfield
   {journal} {\bibinfo  {journal} {Nucl. Phys. B}\ }\textbf {\bibinfo {volume}
  {282}},\ \bibinfo {pages} {163} (\bibinfo {year} {1987})}\BibitemShut
  {NoStop}%
\bibitem [{\citenamefont {Barvinsky}\ and\ \citenamefont
  {Vilkovisky}(1990{\natexlab{a}})}]{Barvinsky:1990up}%
  \BibitemOpen
  \bibfield  {author} {\bibinfo {author} {\bibfnamefont {A.~O.}\ \bibnamefont
  {Barvinsky}}\ and\ \bibinfo {author} {\bibfnamefont {G.~A.}\ \bibnamefont
  {Vilkovisky}},\ }\bibfield  {title} {\bibinfo {title} {{Covariant
  perturbation theory. 2: Second order in the curvature. General algorithms}},\
  }\href {https://doi.org/10.1016/0550-3213(90)90047-H} {\bibfield  {journal}
  {\bibinfo  {journal} {Nucl. Phys. B}\ }\textbf {\bibinfo {volume} {333}},\
  \bibinfo {pages} {471} (\bibinfo {year} {1990}{\natexlab{a}})}\BibitemShut
  {NoStop}%
\bibitem [{\citenamefont {Schwinger}(1951)}]{Schwinger:1951nm}%
  \BibitemOpen
  \bibfield  {author} {\bibinfo {author} {\bibfnamefont {J.~S.}\ \bibnamefont
  {Schwinger}},\ }\bibfield  {title} {\bibinfo {title} {{On gauge invariance
  and vacuum polarization}},\ }\href {https://doi.org/10.1103/PhysRev.82.664}
  {\bibfield  {journal} {\bibinfo  {journal} {Phys. Rev.}\ }\textbf {\bibinfo
  {volume} {82}},\ \bibinfo {pages} {664} (\bibinfo {year} {1951})}\BibitemShut
  {NoStop}%
\bibitem [{\citenamefont {DeWitt}(1965)}]{DeWitt:1964mxt}%
  \BibitemOpen
  \bibfield  {author} {\bibinfo {author} {\bibfnamefont {B.~S.}\ \bibnamefont
  {DeWitt}},\ }\href@noop {} {\emph {\bibinfo {title} {Dynamical Theory of
  Groups and Fields}}},\ Documents on Modern Physics\ (\bibinfo  {publisher}
  {Gordon and Breach},\ \bibinfo {address} {New York},\ \bibinfo {year}
  {1965})\BibitemShut {NoStop}%
\bibitem [{\citenamefont {Barvinsky}\ and\ \citenamefont
  {Vilkovisky}(1985)}]{Barvinsky:1985an}%
  \BibitemOpen
  \bibfield  {author} {\bibinfo {author} {\bibfnamefont {A.~O.}\ \bibnamefont
  {Barvinsky}}\ and\ \bibinfo {author} {\bibfnamefont {G.~A.}\ \bibnamefont
  {Vilkovisky}},\ }\bibfield  {title} {\bibinfo {title} {{The Generalized
  Schwinger-Dewitt Technique in Gauge Theories and Quantum Gravity}},\ }\href
  {https://doi.org/10.1016/0370-1573(85)90148-6} {\bibfield  {journal}
  {\bibinfo  {journal} {Phys. Rept.}\ }\textbf {\bibinfo {volume} {119}},\
  \bibinfo {pages} {1} (\bibinfo {year} {1985})}\BibitemShut {NoStop}%
\bibitem [{\citenamefont {Seeley}(1967)}]{Seeley:1967ea}%
  \BibitemOpen
  \bibfield  {author} {\bibinfo {author} {\bibfnamefont {R.~T.}\ \bibnamefont
  {Seeley}},\ }\bibfield  {title} {\bibinfo {title} {{Complex powers of an
  elliptic operator}},\ }\href@noop {} {\bibfield  {journal} {\bibinfo
  {journal} {Proc. Symp. Pure Math.}\ }\textbf {\bibinfo {volume} {10}},\
  \bibinfo {pages} {288} (\bibinfo {year} {1967})}\BibitemShut {NoStop}%
\bibitem [{\citenamefont {Gilkey}(1975)}]{Gilkey:1975iq}%
  \BibitemOpen
  \bibfield  {author} {\bibinfo {author} {\bibfnamefont {P.~B.}\ \bibnamefont
  {Gilkey}},\ }\bibfield  {title} {\bibinfo {title} {{The Spectral geometry of
  a Riemannian manifold}},\ }\href {https://doi.org/10.4310/jdg/1214433164}
  {\bibfield  {journal} {\bibinfo  {journal} {J. Diff. Geom.}\ }\textbf
  {\bibinfo {volume} {10}},\ \bibinfo {pages} {601} (\bibinfo {year}
  {1975})}\BibitemShut {NoStop}%
\bibitem [{\citenamefont {Deser}\ \emph {et~al.}(1976)\citenamefont {Deser},
  \citenamefont {Duff},\ and\ \citenamefont {Isham}}]{Deser:1976yx}%
  \BibitemOpen
  \bibfield  {author} {\bibinfo {author} {\bibfnamefont {S.}~\bibnamefont
  {Deser}}, \bibinfo {author} {\bibfnamefont {M.~J.}\ \bibnamefont {Duff}},\
  and\ \bibinfo {author} {\bibfnamefont {C.~J.}\ \bibnamefont {Isham}},\
  }\bibfield  {title} {\bibinfo {title} {{Nonlocal Conformal Anomalies}},\
  }\href {https://doi.org/10.1016/0550-3213(76)90480-6} {\bibfield  {journal}
  {\bibinfo  {journal} {Nucl. Phys. B}\ }\textbf {\bibinfo {volume} {111}},\
  \bibinfo {pages} {45} (\bibinfo {year} {1976})}\BibitemShut {NoStop}%
\bibitem [{\citenamefont {Barvinsky}\ \emph {et~al.}(1993)\citenamefont
  {Barvinsky}, \citenamefont {Gusev}, \citenamefont {Zhytnikov},\ and\
  \citenamefont {Vilkovisky}}]{Barvinsky:1993en}%
  \BibitemOpen
  \bibfield  {author} {\bibinfo {author} {\bibfnamefont {A.~O.}\ \bibnamefont
  {Barvinsky}}, \bibinfo {author} {\bibfnamefont {Y.~V.}\ \bibnamefont
  {Gusev}}, \bibinfo {author} {\bibfnamefont {V.~V.}\ \bibnamefont
  {Zhytnikov}},\ and\ \bibinfo {author} {\bibfnamefont {G.~A.}\ \bibnamefont
  {Vilkovisky}},\ }\bibfield  {title} {\bibinfo {title} {{Covariant
  perturbation theory. 4. Third order in the curvature}},\ }\href@noop {} {\
  (\bibinfo {year} {1993})},\ \Eprint {https://arxiv.org/abs/0911.1168}
  {arXiv:0911.1168 [hep-th]} \BibitemShut {NoStop}%
\bibitem [{\citenamefont {Barvinsky}\ \emph {et~al.}(2022)\citenamefont
  {Barvinsky}, \citenamefont {Kurov},\ and\ \citenamefont
  {Sibiryakov}}]{Barvinsky:2021ubv}%
  \BibitemOpen
  \bibfield  {author} {\bibinfo {author} {\bibfnamefont {A.~O.}\ \bibnamefont
  {Barvinsky}}, \bibinfo {author} {\bibfnamefont {A.~V.}\ \bibnamefont
  {Kurov}},\ and\ \bibinfo {author} {\bibfnamefont {S.~M.}\ \bibnamefont
  {Sibiryakov}},\ }\bibfield  {title} {\bibinfo {title} {{Beta functions of
  (3+1)-dimensional projectable Ho{\v{r}}ava gravity}},\ }\href
  {https://doi.org/10.1103/PhysRevD.105.044009} {\bibfield  {journal} {\bibinfo
   {journal} {Phys. Rev. D}\ }\textbf {\bibinfo {volume} {105}},\ \bibinfo
  {pages} {044009} (\bibinfo {year} {2022})},\ \Eprint
  {https://arxiv.org/abs/2110.14688} {arXiv:2110.14688 [hep-th]} \BibitemShut
  {NoStop}%
\bibitem [{\citenamefont {Donoghue}\ and\ \citenamefont
  {El-Menoufi}(2014)}]{Donoghue:2014yha}%
  \BibitemOpen
  \bibfield  {author} {\bibinfo {author} {\bibfnamefont {J.~F.}\ \bibnamefont
  {Donoghue}}\ and\ \bibinfo {author} {\bibfnamefont {B.~K.}\ \bibnamefont
  {El-Menoufi}},\ }\bibfield  {title} {\bibinfo {title} {{Nonlocal quantum
  effects in cosmology: Quantum memory, nonlocal FLRW equations, and
  singularity avoidance}},\ }\href {https://doi.org/10.1103/PhysRevD.89.104062}
  {\bibfield  {journal} {\bibinfo  {journal} {Phys. Rev. D}\ }\textbf {\bibinfo
  {volume} {89}},\ \bibinfo {pages} {104062} (\bibinfo {year} {2014})},\
  \Eprint {https://arxiv.org/abs/1402.3252} {arXiv:1402.3252 [gr-qc]}
  \BibitemShut {NoStop}%
\bibitem [{\citenamefont {Donoghue}\ and\ \citenamefont
  {El-Menoufi}(2015{\natexlab{a}})}]{Donoghue:2015nba}%
  \BibitemOpen
  \bibfield  {author} {\bibinfo {author} {\bibfnamefont {J.~F.}\ \bibnamefont
  {Donoghue}}\ and\ \bibinfo {author} {\bibfnamefont {B.~K.}\ \bibnamefont
  {El-Menoufi}},\ }\bibfield  {title} {\bibinfo {title} {{Covariant non-local
  action for massless QED and the curvature expansion}},\ }\href
  {https://doi.org/10.1007/JHEP10(2015)044} {\bibfield  {journal} {\bibinfo
  {journal} {JHEP}\ }\textbf {\bibinfo {volume} {10}},\ \bibinfo {pages}
  {044}},\ \Eprint {https://arxiv.org/abs/1507.06321} {arXiv:1507.06321
  [hep-th]} \BibitemShut {NoStop}%
\bibitem [{\citenamefont {Donoghue}\ \emph {et~al.}(2017)\citenamefont
  {Donoghue}, \citenamefont {Ivanov},\ and\ \citenamefont
  {Shkerin}}]{Donoghue:2017pgk}%
  \BibitemOpen
  \bibfield  {author} {\bibinfo {author} {\bibfnamefont {J.~F.}\ \bibnamefont
  {Donoghue}}, \bibinfo {author} {\bibfnamefont {M.~M.}\ \bibnamefont
  {Ivanov}},\ and\ \bibinfo {author} {\bibfnamefont {A.}~\bibnamefont
  {Shkerin}},\ }\bibfield  {title} {\bibinfo {title} {{EPFL Lectures on General
  Relativity as a Quantum Field Theory}},\ }\href@noop {} {\  (\bibinfo {year}
  {2017})},\ \Eprint {https://arxiv.org/abs/1702.00319} {arXiv:1702.00319
  [hep-th]} \BibitemShut {NoStop}%
\bibitem [{\citenamefont {Adshead}\ \emph {et~al.}(2009)\citenamefont
  {Adshead}, \citenamefont {Easther},\ and\ \citenamefont
  {Lim}}]{Adshead:2009cb}%
  \BibitemOpen
  \bibfield  {author} {\bibinfo {author} {\bibfnamefont {P.}~\bibnamefont
  {Adshead}}, \bibinfo {author} {\bibfnamefont {R.}~\bibnamefont {Easther}},\
  and\ \bibinfo {author} {\bibfnamefont {E.~A.}\ \bibnamefont {Lim}},\
  }\bibfield  {title} {\bibinfo {title} {{The 'in-in' Formalism and
  Cosmological Perturbations}},\ }\href
  {https://doi.org/10.1103/PhysRevD.80.083521} {\bibfield  {journal} {\bibinfo
  {journal} {Phys. Rev. D}\ }\textbf {\bibinfo {volume} {80}},\ \bibinfo
  {pages} {083521} (\bibinfo {year} {2009})},\ \Eprint
  {https://arxiv.org/abs/0904.4207} {arXiv:0904.4207 [hep-th]} \BibitemShut
  {NoStop}%
\bibitem [{\citenamefont {Schwinger}(1961)}]{Schwinger:1960qe}%
  \BibitemOpen
  \bibfield  {author} {\bibinfo {author} {\bibfnamefont {J.~S.}\ \bibnamefont
  {Schwinger}},\ }\bibfield  {title} {\bibinfo {title} {{Brownian motion of a
  quantum oscillator}},\ }\href {https://doi.org/10.1063/1.1703727} {\bibfield
  {journal} {\bibinfo  {journal} {J. Math. Phys.}\ }\textbf {\bibinfo {volume}
  {2}},\ \bibinfo {pages} {407} (\bibinfo {year} {1961})}\BibitemShut {NoStop}%
\bibitem [{\citenamefont {Keldysh}(1965)}]{Keldysh:1964ud}%
  \BibitemOpen
  \bibfield  {author} {\bibinfo {author} {\bibfnamefont {L.~V.}\ \bibnamefont
  {Keldysh}},\ }\bibfield  {title} {\bibinfo {title} {{Diagram Technique for
  Nonequilibrium Processes}},\ }\href
  {https://doi.org/10.1142/9789811279461_0007} {\bibfield  {journal} {\bibinfo
  {journal} {Sov. Phys. JETP}\ }\textbf {\bibinfo {volume} {20}},\ \bibinfo
  {pages} {1018} (\bibinfo {year} {1965})}\BibitemShut {NoStop}%
\bibitem [{\citenamefont {Jordan}(1986)}]{Jordan:1986ug}%
  \BibitemOpen
  \bibfield  {author} {\bibinfo {author} {\bibfnamefont {R.~D.}\ \bibnamefont
  {Jordan}},\ }\bibfield  {title} {\bibinfo {title} {{Effective Field Equations
  for Expectation Values}},\ }\href {https://doi.org/10.1103/PhysRevD.33.444}
  {\bibfield  {journal} {\bibinfo  {journal} {Phys. Rev. D}\ }\textbf {\bibinfo
  {volume} {33}},\ \bibinfo {pages} {444} (\bibinfo {year} {1986})}\BibitemShut
  {NoStop}%
\bibitem [{\citenamefont {Calzetta}\ and\ \citenamefont
  {Hu}(1987)}]{Calzetta:1986ey}%
  \BibitemOpen
  \bibfield  {author} {\bibinfo {author} {\bibfnamefont {E.}~\bibnamefont
  {Calzetta}}\ and\ \bibinfo {author} {\bibfnamefont {B.~L.}\ \bibnamefont
  {Hu}},\ }\bibfield  {title} {\bibinfo {title} {{Closed Time Path Functional
  Formalism in Curved Space-Time: Application to Cosmological Back Reaction
  Problems}},\ }\href {https://doi.org/10.1103/PhysRevD.35.495} {\bibfield
  {journal} {\bibinfo  {journal} {Phys. Rev. D}\ }\textbf {\bibinfo {volume}
  {35}},\ \bibinfo {pages} {495} (\bibinfo {year} {1987})}\BibitemShut
  {NoStop}%
\bibitem [{\citenamefont {Ford}\ and\ \citenamefont
  {Woodard}(2005)}]{Ford:2004wc}%
  \BibitemOpen
  \bibfield  {author} {\bibinfo {author} {\bibfnamefont {L.~H.}\ \bibnamefont
  {Ford}}\ and\ \bibinfo {author} {\bibfnamefont {R.~P.}\ \bibnamefont
  {Woodard}},\ }\bibfield  {title} {\bibinfo {title} {{Stress tensor
  correlators in the Schwinger-Keldysh formalism}},\ }\href
  {https://doi.org/10.1088/0264-9381/22/9/011} {\bibfield  {journal} {\bibinfo
  {journal} {Class. Quant. Grav.}\ }\textbf {\bibinfo {volume} {22}},\ \bibinfo
  {pages} {1637} (\bibinfo {year} {2005})},\ \Eprint
  {https://arxiv.org/abs/gr-qc/0411003} {arXiv:gr-qc/0411003} \BibitemShut
  {NoStop}%
\bibitem [{\citenamefont {Onemli}\ and\ \citenamefont
  {Woodard}(2002)}]{Onemli:2002hr}%
  \BibitemOpen
  \bibfield  {author} {\bibinfo {author} {\bibfnamefont {V.~K.}\ \bibnamefont
  {Onemli}}\ and\ \bibinfo {author} {\bibfnamefont {R.~P.}\ \bibnamefont
  {Woodard}},\ }\bibfield  {title} {\bibinfo {title} {{Superacceleration from
  massless, minimally coupled phi**4}},\ }\href
  {https://doi.org/10.1088/0264-9381/19/17/311} {\bibfield  {journal} {\bibinfo
   {journal} {Class. Quant. Grav.}\ }\textbf {\bibinfo {volume} {19}},\
  \bibinfo {pages} {4607} (\bibinfo {year} {2002})},\ \Eprint
  {https://arxiv.org/abs/gr-qc/0204065} {arXiv:gr-qc/0204065} \BibitemShut
  {NoStop}%
\bibitem [{\citenamefont {Barvinsky}\ and\ \citenamefont
  {Kolganov}(2024)}]{Barvinsky:2023jkl}%
  \BibitemOpen
  \bibfield  {author} {\bibinfo {author} {\bibfnamefont {A.~O.}\ \bibnamefont
  {Barvinsky}}\ and\ \bibinfo {author} {\bibfnamefont {N.}~\bibnamefont
  {Kolganov}},\ }\bibfield  {title} {\bibinfo {title} {{Nonequilibrium
  Schwinger-Keldysh formalism for density matrix states: Analytic properties
  and implications in cosmology}},\ }\href
  {https://doi.org/10.1103/PhysRevD.109.025004} {\bibfield  {journal} {\bibinfo
   {journal} {Phys. Rev. D}\ }\textbf {\bibinfo {volume} {109}},\ \bibinfo
  {pages} {025004} (\bibinfo {year} {2024})},\ \Eprint
  {https://arxiv.org/abs/2309.03687} {arXiv:2309.03687 [hep-th]} \BibitemShut
  {NoStop}%
\bibitem [{\citenamefont {Dowker}\ and\ \citenamefont
  {Schofield}(1988)}]{Dowker:1988jw}%
  \BibitemOpen
  \bibfield  {author} {\bibinfo {author} {\bibfnamefont {J.~S.}\ \bibnamefont
  {Dowker}}\ and\ \bibinfo {author} {\bibfnamefont {J.~P.}\ \bibnamefont
  {Schofield}},\ }\bibfield  {title} {\bibinfo {title} {{High Temperature
  Expansion of the Free Energy of a Massive Scalar Field in a Curved Space}},\
  }\href {https://doi.org/10.1103/PhysRevD.38.3327} {\bibfield  {journal}
  {\bibinfo  {journal} {Phys. Rev. D}\ }\textbf {\bibinfo {volume} {38}},\
  \bibinfo {pages} {3327} (\bibinfo {year} {1988})}\BibitemShut {NoStop}%
\bibitem [{\citenamefont {Gusev}\ and\ \citenamefont
  {Zelnikov}(1999)}]{Gusev:1998rp}%
  \BibitemOpen
  \bibfield  {author} {\bibinfo {author} {\bibfnamefont {Y.~V.}\ \bibnamefont
  {Gusev}}\ and\ \bibinfo {author} {\bibfnamefont {A.~I.}\ \bibnamefont
  {Zelnikov}},\ }\bibfield  {title} {\bibinfo {title} {{Finite temperature
  nonlocal effective action for quantum fields in curved space}},\ }\href
  {https://doi.org/10.1103/PhysRevD.59.024002} {\bibfield  {journal} {\bibinfo
  {journal} {Phys. Rev. D}\ }\textbf {\bibinfo {volume} {59}},\ \bibinfo
  {pages} {024002} (\bibinfo {year} {1999})},\ \Eprint
  {https://arxiv.org/abs/hep-th/9807038} {arXiv:hep-th/9807038} \BibitemShut
  {NoStop}%
\bibitem [{\citenamefont {El{\'\i}as}\ \emph {et~al.}(2017)\citenamefont
  {El{\'\i}as}, \citenamefont {Mazzitelli},\ and\ \citenamefont
  {Trombetta}}]{Elias:2017wkr}%
  \BibitemOpen
  \bibfield  {author} {\bibinfo {author} {\bibfnamefont {M.}~\bibnamefont
  {El{\'\i}as}}, \bibinfo {author} {\bibfnamefont {F.~D.}\ \bibnamefont
  {Mazzitelli}},\ and\ \bibinfo {author} {\bibfnamefont {L.~G.}\ \bibnamefont
  {Trombetta}},\ }\bibfield  {title} {\bibinfo {title} {{Nonlocal effective
  actions in semiclassical gravity: thermal effects in stationary
  geometries}},\ }\href {https://doi.org/10.1103/PhysRevD.96.105007} {\bibfield
   {journal} {\bibinfo  {journal} {Phys. Rev. D}\ }\textbf {\bibinfo {volume}
  {96}},\ \bibinfo {pages} {105007} (\bibinfo {year} {2017})},\ \Eprint
  {https://arxiv.org/abs/1709.10435} {arXiv:1709.10435 [hep-th]} \BibitemShut
  {NoStop}%
\bibitem [{\citenamefont {Valle}\ and\ \citenamefont
  {Vazquez-Mozo}(2025)}]{Valle:2025fev}%
  \BibitemOpen
  \bibfield  {author} {\bibinfo {author} {\bibfnamefont {M.}~\bibnamefont
  {Valle}}\ and\ \bibinfo {author} {\bibfnamefont {M.~A.}\ \bibnamefont
  {Vazquez-Mozo}},\ }\bibfield  {title} {\bibinfo {title} {{High-temperature
  massless scalar partition function on general stationary~backgrounds}},\
  }\href {https://doi.org/10.1103/m5qv-q8hq} {\bibfield  {journal} {\bibinfo
  {journal} {Phys. Rev. D}\ }\textbf {\bibinfo {volume} {112}},\ \bibinfo
  {pages} {105008} (\bibinfo {year} {2025})},\ \Eprint
  {https://arxiv.org/abs/2509.01333} {arXiv:2509.01333 [hep-th]} \BibitemShut
  {NoStop}%
\bibitem [{\citenamefont {Donoghue}\ and\ \citenamefont
  {El-Menoufi}(2015{\natexlab{b}})}]{Donoghue:2015xla}%
  \BibitemOpen
  \bibfield  {author} {\bibinfo {author} {\bibfnamefont {J.~F.}\ \bibnamefont
  {Donoghue}}\ and\ \bibinfo {author} {\bibfnamefont {B.~K.}\ \bibnamefont
  {El-Menoufi}},\ }\bibfield  {title} {\bibinfo {title} {{QED trace anomaly,
  non-local Lagrangians and quantum Equivalence Principle violations}},\ }\href
  {https://doi.org/10.1007/JHEP05(2015)118} {\bibfield  {journal} {\bibinfo
  {journal} {JHEP}\ }\textbf {\bibinfo {volume} {05}},\ \bibinfo {pages}
  {118}},\ \Eprint {https://arxiv.org/abs/1503.06099} {arXiv:1503.06099
  [hep-th]} \BibitemShut {NoStop}%
\bibitem [{\citenamefont {Codello}\ and\ \citenamefont
  {Zanusso}(2013)}]{Codello:2012kq}%
  \BibitemOpen
  \bibfield  {author} {\bibinfo {author} {\bibfnamefont {A.}~\bibnamefont
  {Codello}}\ and\ \bibinfo {author} {\bibfnamefont {O.}~\bibnamefont
  {Zanusso}},\ }\bibfield  {title} {\bibinfo {title} {{On the non-local heat
  kernel expansion}},\ }\href {https://doi.org/10.1063/1.4776234} {\bibfield
  {journal} {\bibinfo  {journal} {J. Math. Phys.}\ }\textbf {\bibinfo {volume}
  {54}},\ \bibinfo {pages} {013513} (\bibinfo {year} {2013})},\ \Eprint
  {https://arxiv.org/abs/1203.2034} {arXiv:1203.2034 [math-ph]} \BibitemShut
  {NoStop}%
\bibitem [{\citenamefont {Vassilevich}(2003)}]{Vassilevich:2003xt}%
  \BibitemOpen
  \bibfield  {author} {\bibinfo {author} {\bibfnamefont {D.~V.}\ \bibnamefont
  {Vassilevich}},\ }\bibfield  {title} {\bibinfo {title} {{Heat kernel
  expansion: User's manual}},\ }\href
  {https://doi.org/10.1016/j.physrep.2003.09.002} {\bibfield  {journal}
  {\bibinfo  {journal} {Phys. Rept.}\ }\textbf {\bibinfo {volume} {388}},\
  \bibinfo {pages} {279} (\bibinfo {year} {2003})},\ \Eprint
  {https://arxiv.org/abs/hep-th/0306138} {arXiv:hep-th/0306138} \BibitemShut
  {NoStop}%
\bibitem [{\citenamefont {Barvinsky}\ and\ \citenamefont {Kolganov}()}]{tbp}%
  \BibitemOpen
  \bibfield  {author} {\bibinfo {author} {\bibfnamefont {A.~O.}\ \bibnamefont
  {Barvinsky}}\ and\ \bibinfo {author} {\bibfnamefont {N.}~\bibnamefont
  {Kolganov}},\ }\href@noop {} {}\bibinfo {howpublished} {to be
  published}\BibitemShut {NoStop}%
\bibitem [{\citenamefont {Barvinsky}\ and\ \citenamefont
  {Vilkovisky}(1990{\natexlab{b}})}]{Barvinsky:1990uq}%
  \BibitemOpen
  \bibfield  {author} {\bibinfo {author} {\bibfnamefont {A.~O.}\ \bibnamefont
  {Barvinsky}}\ and\ \bibinfo {author} {\bibfnamefont {G.~A.}\ \bibnamefont
  {Vilkovisky}},\ }\bibfield  {title} {\bibinfo {title} {{Covariant
  perturbation theory. 3: Spectral representations of the third order
  form-factors}},\ }\href {https://doi.org/10.1016/0550-3213(90)90048-I}
  {\bibfield  {journal} {\bibinfo  {journal} {Nucl. Phys. B}\ }\textbf
  {\bibinfo {volume} {333}},\ \bibinfo {pages} {512} (\bibinfo {year}
  {1990}{\natexlab{b}})}\BibitemShut {NoStop}%
\bibitem [{\citenamefont {Vilkovisky}(2008)}]{Vilkovisky:2007ny}%
  \BibitemOpen
  \bibfield  {author} {\bibinfo {author} {\bibfnamefont {G.~A.}\ \bibnamefont
  {Vilkovisky}},\ }\bibfield  {title} {\bibinfo {title} {{Expectation values
  and vacuum currents of quantum fields}},\ }\href@noop {} {\bibfield
  {journal} {\bibinfo  {journal} {Lect. Notes Phys.}\ }\textbf {\bibinfo
  {volume} {737}},\ \bibinfo {pages} {729} (\bibinfo {year} {2008})},\ \Eprint
  {https://arxiv.org/abs/0712.3379} {arXiv:0712.3379 [hep-th]} \BibitemShut
  {NoStop}%
\bibitem [{\citenamefont {Barvinsky}(2007)}]{Barvinsky:2007vb}%
  \BibitemOpen
  \bibfield  {author} {\bibinfo {author} {\bibfnamefont {A.~O.}\ \bibnamefont
  {Barvinsky}},\ }\bibfield  {title} {\bibinfo {title} {{Why there is something
  rather than nothing (out of everything)?}},\ }\href
  {https://doi.org/10.1103/PhysRevLett.99.071301} {\bibfield  {journal}
  {\bibinfo  {journal} {Phys. Rev. Lett.}\ }\textbf {\bibinfo {volume} {99}},\
  \bibinfo {pages} {071301} (\bibinfo {year} {2007})},\ \Eprint
  {https://arxiv.org/abs/0704.0083} {arXiv:0704.0083 [hep-th]} \BibitemShut
  {NoStop}%
\bibitem [{\citenamefont {Barvinsky}\ and\ \citenamefont
  {Kamenshchik}(2006)}]{Barvinsky:2006uh}%
  \BibitemOpen
  \bibfield  {author} {\bibinfo {author} {\bibfnamefont {A.~O.}\ \bibnamefont
  {Barvinsky}}\ and\ \bibinfo {author} {\bibfnamefont {A.~Y.}\ \bibnamefont
  {Kamenshchik}},\ }\bibfield  {title} {\bibinfo {title} {{Cosmological
  landscape from nothing: Some like it hot}},\ }\href
  {https://doi.org/10.1088/1475-7516/2006/09/014} {\bibfield  {journal}
  {\bibinfo  {journal} {JCAP}\ }\textbf {\bibinfo {volume} {09}},\ \bibinfo
  {pages} {014}},\ \Eprint {https://arxiv.org/abs/hep-th/0605132}
  {arXiv:hep-th/0605132} \BibitemShut {NoStop}%
\bibitem [{\citenamefont {Starobinsky}(1980)}]{Starobinsky:1980te}%
  \BibitemOpen
  \bibfield  {author} {\bibinfo {author} {\bibfnamefont {A.~A.}\ \bibnamefont
  {Starobinsky}},\ }\bibfield  {title} {\bibinfo {title} {{A New Type of
  Isotropic Cosmological Models Without Singularity}},\ }\href
  {https://doi.org/10.1016/0370-2693(80)90670-X} {\bibfield  {journal}
  {\bibinfo  {journal} {Phys. Lett. B}\ }\textbf {\bibinfo {volume} {91}},\
  \bibinfo {pages} {99} (\bibinfo {year} {1980})}\BibitemShut {NoStop}%
\bibitem [{\citenamefont {Barvinsky}\ \emph {et~al.}(2015)\citenamefont
  {Barvinsky}, \citenamefont {Kamenshchik},\ and\ \citenamefont
  {Nesterov}}]{Barvinsky:2015uxa}%
  \BibitemOpen
  \bibfield  {author} {\bibinfo {author} {\bibfnamefont {A.~O.}\ \bibnamefont
  {Barvinsky}}, \bibinfo {author} {\bibfnamefont {A.~Y.}\ \bibnamefont
  {Kamenshchik}},\ and\ \bibinfo {author} {\bibfnamefont {D.~V.}\ \bibnamefont
  {Nesterov}},\ }\bibfield  {title} {\bibinfo {title} {{Origin of inflation in
  CFT driven cosmology: $R^2$-gravity and non-minimally coupled inflaton
  models}},\ }\href {https://doi.org/10.1140/epjc/s10052-015-3817-7} {\bibfield
   {journal} {\bibinfo  {journal} {Eur. Phys. J. C}\ }\textbf {\bibinfo
  {volume} {75}},\ \bibinfo {pages} {584} (\bibinfo {year} {2015})},\ \Eprint
  {https://arxiv.org/abs/1510.06858} {arXiv:1510.06858 [hep-th]} \BibitemShut
  {NoStop}%
\bibitem [{\citenamefont {Barvinsky}(2016)}]{Barvinsky:2015wvz}%
  \BibitemOpen
  \bibfield  {author} {\bibinfo {author} {\bibfnamefont {A.~O.}\ \bibnamefont
  {Barvinsky}},\ }\bibfield  {title} {\bibinfo {title} {{CFT driven cosmology
  and conformal higher spin fields}},\ }\href
  {https://doi.org/10.1103/PhysRevD.93.103530} {\bibfield  {journal} {\bibinfo
  {journal} {Phys. Rev. D}\ }\textbf {\bibinfo {volume} {93}},\ \bibinfo
  {pages} {103530} (\bibinfo {year} {2016})},\ \Eprint
  {https://arxiv.org/abs/1511.07625} {arXiv:1511.07625 [hep-th]} \BibitemShut
  {NoStop}%
\bibitem [{\citenamefont {Buoninfante}\ \emph {et~al.}(2018)\citenamefont
  {Buoninfante}, \citenamefont {Koshelev}, \citenamefont {Lambiase},\ and\
  \citenamefont {Mazumdar}}]{Buoninfante:2018xiw}%
  \BibitemOpen
  \bibfield  {author} {\bibinfo {author} {\bibfnamefont {L.}~\bibnamefont
  {Buoninfante}}, \bibinfo {author} {\bibfnamefont {A.~S.}\ \bibnamefont
  {Koshelev}}, \bibinfo {author} {\bibfnamefont {G.}~\bibnamefont {Lambiase}},\
  and\ \bibinfo {author} {\bibfnamefont {A.}~\bibnamefont {Mazumdar}},\
  }\bibfield  {title} {\bibinfo {title} {{Classical properties of non-local,
  ghost- and singularity-free gravity}},\ }\href
  {https://doi.org/10.1088/1475-7516/2018/09/034} {\bibfield  {journal}
  {\bibinfo  {journal} {JCAP}\ }\textbf {\bibinfo {volume} {09}},\ \bibinfo
  {pages} {034}},\ \Eprint {https://arxiv.org/abs/1802.00399} {arXiv:1802.00399
  [gr-qc]} \BibitemShut {NoStop}%
\bibitem [{\citenamefont {Koshelev}\ \emph {et~al.}(2018)\citenamefont
  {Koshelev}, \citenamefont {Sravan~Kumar}, \citenamefont {Modesto},\ and\
  \citenamefont {Rachwa{\l}}}]{Koshelev:2017ebj}%
  \BibitemOpen
  \bibfield  {author} {\bibinfo {author} {\bibfnamefont {A.~S.}\ \bibnamefont
  {Koshelev}}, \bibinfo {author} {\bibfnamefont {K.}~\bibnamefont
  {Sravan~Kumar}}, \bibinfo {author} {\bibfnamefont {L.}~\bibnamefont
  {Modesto}},\ and\ \bibinfo {author} {\bibfnamefont {L.}~\bibnamefont
  {Rachwa{\l}}},\ }\bibfield  {title} {\bibinfo {title} {{Finite quantum
  gravity in dS and AdS spacetimes}},\ }\href
  {https://doi.org/10.1103/PhysRevD.98.046007} {\bibfield  {journal} {\bibinfo
  {journal} {Phys. Rev. D}\ }\textbf {\bibinfo {volume} {98}},\ \bibinfo
  {pages} {046007} (\bibinfo {year} {2018})},\ \Eprint
  {https://arxiv.org/abs/1710.07759} {arXiv:1710.07759 [hep-th]} \BibitemShut
  {NoStop}%
\bibitem [{\citenamefont {Koshelev}\ \emph {et~al.}(2025)\citenamefont
  {Koshelev}, \citenamefont {Melichev},\ and\ \citenamefont
  {Rachwal}}]{Koshelev:2025pxg}%
  \BibitemOpen
  \bibfield  {author} {\bibinfo {author} {\bibfnamefont {A.~S.}\ \bibnamefont
  {Koshelev}}, \bibinfo {author} {\bibfnamefont {O.}~\bibnamefont {Melichev}},\
  and\ \bibinfo {author} {\bibfnamefont {L.}~\bibnamefont {Rachwal}},\
  }\bibfield  {title} {\bibinfo {title} {{Cancellation of UV divergences in
  ghost-free infinite derivative gravity}},\ }\href@noop {} {\  (\bibinfo
  {year} {2025})},\ \Eprint {https://arxiv.org/abs/2512.18006}
  {arXiv:2512.18006 [hep-th]} \BibitemShut {NoStop}%
\bibitem [{\citenamefont {Witten}(2026)}]{Witten:2025ayw}%
  \BibitemOpen
  \bibfield  {author} {\bibinfo {author} {\bibfnamefont {E.}~\bibnamefont
  {Witten}},\ }\bibfield  {title} {\bibinfo {title} {{Bras and kets in
  Euclidean path integrals}},\ }\href
  {https://doi.org/10.4310/bpam.260113013520} {\bibfield  {journal} {\bibinfo
  {journal} {Beijing J. Pure Appl. Math.}\ }\textbf {\bibinfo {volume} {3}},\
  \bibinfo {pages} {1} (\bibinfo {year} {2026})},\ \Eprint
  {https://arxiv.org/abs/2503.12771} {arXiv:2503.12771 [hep-th]} \BibitemShut
  {NoStop}%
\bibitem [{\citenamefont {Fumagalli}\ \emph {et~al.}(2025)\citenamefont
  {Fumagalli}, \citenamefont {Gorbenko},\ and\ \citenamefont
  {Kames-King}}]{Fumagalli:2024msi}%
  \BibitemOpen
  \bibfield  {author} {\bibinfo {author} {\bibfnamefont {A.}~\bibnamefont
  {Fumagalli}}, \bibinfo {author} {\bibfnamefont {V.}~\bibnamefont
  {Gorbenko}},\ and\ \bibinfo {author} {\bibfnamefont {J.}~\bibnamefont
  {Kames-King}},\ }\bibfield  {title} {\bibinfo {title} {{De Sitter Bra-Ket
  wormholes}},\ }\href {https://doi.org/10.1007/JHEP05(2025)074} {\bibfield
  {journal} {\bibinfo  {journal} {JHEP}\ }\textbf {\bibinfo {volume} {05}},\
  \bibinfo {pages} {074}},\ \Eprint {https://arxiv.org/abs/2408.08351}
  {arXiv:2408.08351 [hep-th]} \BibitemShut {NoStop}%
\bibitem [{\citenamefont {Blommaert}\ \emph {et~al.}(2025)\citenamefont
  {Blommaert}, \citenamefont {Kudler-Flam},\ and\ \citenamefont
  {Urbach}}]{Blommaert:2025bgd}%
  \BibitemOpen
  \bibfield  {author} {\bibinfo {author} {\bibfnamefont {A.}~\bibnamefont
  {Blommaert}}, \bibinfo {author} {\bibfnamefont {J.}~\bibnamefont
  {Kudler-Flam}},\ and\ \bibinfo {author} {\bibfnamefont {E.~Y.}\ \bibnamefont
  {Urbach}},\ }\bibfield  {title} {\bibinfo {title} {{Absolute entropy and the
  observer{\textquoteright}s no-boundary state}},\ }\href
  {https://doi.org/10.1007/JHEP11(2025)113} {\bibfield  {journal} {\bibinfo
  {journal} {JHEP}\ }\textbf {\bibinfo {volume} {11}},\ \bibinfo {pages}
  {113}},\ \Eprint {https://arxiv.org/abs/2505.14771} {arXiv:2505.14771
  [hep-th]} \BibitemShut {NoStop}%
\bibitem [{\citenamefont {Higuchi}\ \emph {et~al.}(2011)\citenamefont
  {Higuchi}, \citenamefont {Marolf},\ and\ \citenamefont
  {Morrison}}]{Higuchi:2010xt}%
  \BibitemOpen
  \bibfield  {author} {\bibinfo {author} {\bibfnamefont {A.}~\bibnamefont
  {Higuchi}}, \bibinfo {author} {\bibfnamefont {D.}~\bibnamefont {Marolf}},\
  and\ \bibinfo {author} {\bibfnamefont {I.~A.}\ \bibnamefont {Morrison}},\
  }\bibfield  {title} {\bibinfo {title} {{On the Equivalence between Euclidean
  and In-In Formalisms in de Sitter QFT}},\ }\href
  {https://doi.org/10.1103/PhysRevD.83.084029} {\bibfield  {journal} {\bibinfo
  {journal} {Phys. Rev. D}\ }\textbf {\bibinfo {volume} {83}},\ \bibinfo
  {pages} {084029} (\bibinfo {year} {2011})},\ \Eprint
  {https://arxiv.org/abs/1012.3415} {arXiv:1012.3415 [gr-qc]} \BibitemShut
  {NoStop}%
\end{thebibliography}%

\end{document}